\documentclass{an_art}
\usepackage{graphicx}
\usepackage{epsf}

\begin{document}

\begin{titlepage}
\setcounter{page}{1}
\headnote{Astron.~Nachr.~321 (2000) 3, 193--206}
\makeheadline

\title{Magnetic deformation of the white dwarf surface structure 
       \quad\quad\quad (PREPRINT)}

\author{{\sc Christian Fendt}, Potsdam, Germany \\
\medskip
{\small Astrophysikalisches Institut Potsdam} \\
\bigskip
{\sc Dainis Dravins}, Lund, Sweden\\
\medskip
{\small Lund Observatory} \\
}

\date{Received 2000 June 26; accepted 2000 July 12} 
\maketitle

\summary
The influence of strong, large-scale magnetic fields on the structure
and temperature distribution in white dwarf atmospheres is investigated.
Magnetic fields may provide an additional component of pressure
support, thus possibly inflating the atmosphere compared to the
non-magnetic case.
Since the magnetic forces are not isotropic, atmospheric
properties may significantly deviate from spherical symmetry.
In this paper the magnetohydrostatic equilibrium is calculated
numerically in the radial direction for either for small deviations
from different assumptions for the poloidal current distribution.
We generally find indication that the scale height of the magnetic
white dwarf
atmosphere enlarges with magnetic field strength and/or poloidal
current strength.
This is in qualitative agreement with recent spectropolarimetric
observations of Grw$+10\degr8247$. 
Quantitatively, we find for e.g. a mean surface poloidal field strength
of 100 MG and a toroidal field strength of 2-10 MG
an increase of scale height by a factor of 10.
This is indicating that already a small deviation from the initial
force-free dipolar magnetic field may lead to observable effects.
We further propose the method of finite elements for the solution of
the two-dimensional magnetohydrostatic equilibrium including radiation
transport in the diffusive approximation.
We present and discuss preliminary solutions, again indicating
on an expansion of the magnetized atmosphere.
END

\keyw 
stars: atmospheres 
stars: fundamental parameters 
stars: magnetic fields 
stars: white dwarfs 
END
	      
\end{titlepage}

\def\pt{P_{\rm tot}}
\def\pmag{P_{\rm mag}}
\def\pr{P_{\rm rad}}
\def\ps{P_{\star}}
\def\te{T_{\rm eff}}
\def\ln{{\rm ln}}
\def\rs{R_{\star}}
\def\ms{M_{\star}}
\def\bs{B_{\star}}
\def\msun{M_{\sun}}
\def\lsun{L_{\sun}}


\section{Motivation}


Magnetic white dwarfs exhibit surface field strengths of the order of
5 - 500\,MG (0.5 - 50 kT) (see e.g. Chanmugam 1992, Landstreet 1992, 
Schmidt \& Smith 1995).
These fields are globally extended fields, distributed over length scales
of several stellar radii.
In general, model calculations of magnetic white dwarf spectral lines
reveal a dipolar
structure of the magnetosphere, sometimes offset from the stellar center,
and, occasionally, superimposed with higher order multipole moments.

The magnetic pressure associated with a white dwarf magnetic field is about
$(\bs^2/8 \pi) = 4\times10^{12}\,(\bs/10\,{\rm MG})^2
\,{\rm dyne}\,{\rm cm}^{-2}$
(or $4\times10^{11}\,(\bs/{\rm kT})^2{\rm Pa})$,
where $\bs$ denotes the mean surface field strength.
The gas pressure $P$ in the non-degenerate stellar envelope varies largely
over the scale height.
In the outermost regions the magnetic pressure exceeds the thermal pressure
by many orders of magnitude.
Hydrodynamic calculations of non-magnetic white dwarf atmospheres reveal
$P \simeq 10^6\,{\rm dyne\,cm}^{-2}$ $(10^5\,{\rm Pa}$) 
at an optical depth $\tau \simeq 1$ (Steffen et al. 1995).
However, concerning the magnetohydrodynamic force-balance in the atmosphere,
also {\em magnetic tension} $(\vec{B}\cdot\nabla)\vec{B}$ has to be taken into
account. 
A dipolar magnetic field, as observed as the main field component in many
white dwarfs, is essentially force-free, i.e. magnetic
pressure and tension forces exactly cancel. 
Therefore, only non force-free field components of a white dwarf
magnetic field will affect the hydrodynamic equilibrium.

The question arises, whether and how such a magnetic field will influence the
global, hydrodynamic structure?
And further, whether and how the radiative equilibrium and the temperature
distribution is affected?
If this is indeed the case, a change in the global temperature distribution
along the stellar surface might be expected, eventually leading to observable
effects (as solar dark spots are connected with an enhanced magnetic field
strength).
The most direct detection of such effects would be a periodic optical
variability of a rotating single magnetic white dwarf.
 
A potential candidate demonstrating such effects is the very hot, 
highly magnetic DA white dwarf RE\,J0317-853.
Barstow et al. (1995) observe an effective temperature of $\sim$ 50000\,K,
a field strength of $\sim$ 340\,MG (34 kT), and optical oscillations
of $\sim 10\%$ with a period of 725\,s.
Another example is PG\,1031+234, where the observed polarization indicates
a strong magnetic spot of an offset dipolar field component 
(Schmidt et al. 1986). 
Due to the two dipole components, the magnetic field strength varies by a
factor of 5 across the stellar surface. 
The observed rotational period is $3^{\rm h}\,24^{\rm m}$.
There is indication also from spectropolarimetric observations that a 
magnetic field affects the radial hydrostatic equilibrium 
(\"Ostreicher et al. 1992, Friedrich et al. 1994).
From the shape of absorption lines in Grw+70\degr8247 these authors
concluded that magnetic white dwarfs may have ``{\it a more expanded 
photosphere}''
(Friedrich et al. 1994).
The corresponding geometrical scale height at the optical depth 
$\tau \simeq 1$ for continuum absorption is increased by a factor of 300.


\subsection{Models of the magnetic structure of white dwarfs}


The fundamental question of the white dwarf magnetic field origin 
and evolution has not yet been settled in the literature
(see Chanmugam 1992 for a review).

The idea that magnetic fields of white dwarfs are generated or maintained
by dynamo action is especially interesting for white dwarfs with variable
magnetic field.
For the example of GD\,358, a white dwarf with only mild field strength 
of 1300 G with variations of $\pm 300 $, Markiel et al. (1994) 
performed dynamo simulations,
indicating that an $\alpha\omega$ dynamo is able to reproduce the
'observed' variation in field strength.
Indeed, Thomas et al. (1995) concluded on theoretical grounds that dynamo
generated magnetic fields could occur in many DB and DA white dwarfs.

As the magnetic flux in Ap stars is comparable to that of magnetic white
dwarfs,
it has been discussed that the magnetic fields in white dwarfs could
just be a relict of (decaying) {\em primordial} stellar fields
(Chanmugam 1992, Landstreet 1992).
For example, Moss (1979b) theoretically investigated axisymmetric dipolar
type magnetic fields in white dwarfs finding that the ratio of internal
to surface magnetic flux increases with the ratio of rotational to
magnetic energy.
Further, convection may prevent a weak magnetic field (below 1\,MG) 
to penetrate the stellar surface.

Most of the literature on the global stellar magnetic field evolution
considers upper main sequence stars and not the white dwarfs.
These papers concentrate on the field evolution and 
only rarely take into account hydrodynamical or thermal properties.
In the following we will briefly review a few examples which are 
more relevant to our work.

Early models were presented by Moss (1975) and Mestel \& Moss (1977)
investigating magnetic upper main sequence stars,
especially the stability of the magnetic field configuration.
Mestel et al. (1988) considered the interaction of magnetic fields,
rotation and meridional circulation in the radiative envelope of
stars.
However, as their aim was the distribution of the differential rotation 
and the toroidal magnetic field,
they prescribed the poloidal components of field and velocity.
Furthermore, they treat a homogeneous sphere of an incompressible 
fluid.

In general, Lorentz forces seem to be able to reduce the effective
gravitational potential in the upper atmosphere by 70-80\%
(Stepie\'n 1978).
Therefore, the whole atmosphere should expand.
Stift (1977, 1978) presented a 
In the Stift (1977, 1978) model of the variable magnetic star 
$\beta\,{\rm CrB}$ it's magnetic variation is explained by a
``possible magnetic deformation'', 
leading to a deviation from spherical symmetry and to precession.
This interpretation, however, is was questioned by Moss (1979a)
on the base of the stellar evolutionary status of that star.
For magnetic Ap stars, Hubbard \& Dearborn (1982) theoretically find the
interesting effect that, due to the additional magnetic pressure, the 
star is ``blown up'' by about $\sim 20\%$, forming a ``magnetic balloon''.

In the case of magnetic white dwarfs, Ostriker \& Hartwick (1968) concluded
that the radius of the degenerate core increases with increasing magnetic
energy in the interior.
For a magnetic energy of 0.1 in units of gravitational energy, the
core radius is 1.4 times larger compared to a non-magnetic core.
In the similar direction goes the recent work by Suh \& Mathews (2000)
who revise the mass-radius relation for magnetic white dwarfs.

Wendell et al. (1987) first presented quantitative results of the global
magnetic field evolution of white dwarfs.
They calculated the time-dependent MHD equations for purely poloidal 
fields neglecting non-radial motion and also any back reaction of the
field on the stellar structure.
In comparison with static models they find that the decay time scale
increases, becoming even longer than the age of the star.
Muslimov et al. (1995) extended this study by including the Hall effect,

The structure of the magnetic white dwarf atmosphere is more extreme than
that of their possible progenitors, the Ap stars.
As the stellar radius is reduced by a large factor and the atmospheric
layer becomes thiner, huge gradients are implied in the hydrodynamic
properties.
In contrary, the magnetic field, if we assume it to have a roughly
dipolar structure, is relatively constant throughout the atmosphere.

The general difficulty in calculating the action of Lorentz forces in
stellar atmospheres is that the electric current distribution is uncertain
(Landstreet 1987).
In the case of  magnetic CP stars observations give no evidence for 
the presence of toroidal fields in their line forming region
(Landstreet 1987).
Although there is not very much known about the distribution of 
toroidal fields below the surface of stars,
electric currents should exist in the stellar interior in order
to maintain a large-scale dipolar field.
Further, Mestel \& Moss (1983) argue that a stellar magnetic field
{\em must} have a mixed poloidal-toroidal structure in order to avoid 
dynamical instabilities.
Outside of the star, in the very low density regions, an axisymmetric 
magnetic field should be essentially force-free.
The question is, whether an electric current system in the stellar 
interior also extends throughout the stellar atmosphere.
In the case of white dwarfs, where the extent of the atmosphere is only
$\sim 1\%$ of the whole star,
we see no reason why the current should be excluded from the surface
sheet
(with 'surface' we denote that layer in the atmosphere, 
where the radiation originates).

Assuming that the magnetic field of Ap stars and white dwarfs is a slowly
decaying fossil magnetic field,
Landstreet (1987) estimated the amount of induction of electric current
from that decay and, correspondingly, the strength of the induced
Lorentz force.
In such a picture the generation of electric current is self-consistently
adopted by a physical process.
However no detailed treatment of the local force-balance was made.
Concerning the white dwarfs, he estimated that the decay-induced Lorentz
force cannot be neglected for the hydrostatic structure of low-mass, hot
white dwarfs.
He argued that on order to affect the hydrostatic structure of the
atmosphere, the Lorentz force should be at least equal in magnitude
compared to the gravitational force.
%
%

Landstreet (1987) further claimed that due to the large Lorentz force a
static equilibrium as discussed by Stepie\'n (1978) is unlikely.
He argued that the star cannot have large un-balanced forces acting in
the outer atmosphere where gravity is unable to balance the Lorentz force.
This would immediately lead to a rapid acceleration.
The decay-induced Lorentz force is normal to the field lines and,
as he proposes, 
tends to drive meridional motion across the field line, 
``probably mainly horizontally, since gas pressure will limit
vertical motion''.
In turn, this would lead to electro-motive forcing currents opposed to
those driven by decay.
The matter will be accelerated
until it is moving in a nearly force-free
configuration.

Our viewpoint is the following.
In the region where large Lorentz forces are present, the frozen-in field
will also guide the plasma motion (see Sect.\,2.2).
Therefore, acceleration in meridional direction across the field is
suppressed by the field.
The plasma will move slowly until a 
{\em new non force-free magnetohydrostatic equilibrium}
is reached.
On the long term evolution magnetic diffusion will nevertheless occur.

So far, no attempt has been undertaken to calculate the magnetohydrostatic
equilibrium in the magnetic white dwarf atmosphere.
Motivated by recent observations of spectropolarimetry, we thought
that this assumption would be a reasonable approach to investigate
the influence of the strong white dwarf magnetic field on their surface
structure.


\subsection{Aim of the present study}


As the previous literature survey has shown, 
a fully self-consistent treatment of the magnetohydro{\em dynamical}
equations is not yet feasible today.
In particular, the combined problem of field generation and field
evolution is far from being solved.

In contrast to the literature on the long term evolution of the
magnetic field in white dwarfs (e.g. Wendell et al. 1987),
we therefore concentrate on a 'snapshot' during this evolution.
For a fixed time, we investigate how the hydrostatic equilibrium would be
affected under the presence of a magnetic field.
We prescribe the magnetic field structure and search for solutions of
the magnetohydrostatic equilibrium in the white dwarf atmosphere.
We will show that such solutions exist.

One may argue that a prescription of some arbitrary toroidal current
distribution is physically not meaningful.
However, it is yet the only way to obtain some quantitative results,
since self-consistent magnetohydrodynamic models of the whole star are
not yet possible to calculate.
We further note that we do not calculate a global stellar model,
but concentrate only on the atmospheric layer of the white dwarf.
A treatment of the MHD of the whole white dwarf is not yet 
numerically feasible.
On the other hand, the field structure (poloidal and toroidal) is
certainly determined by the inner structure of the star, which we
treat as a kind of boundary condition for the atmosphere.
Since the lack of knowledge about the true toroidal field / poloidal
electric current distribution, 
it seems appropriate to us to discuss different examples for the current
distribution in order to study how they affect the hydrostatic structure
of the atmosphere.
 
In the present paper, we consider the case only {\em small deviation}
from a force-free dipolar structure. 
Therefore, in order to study the potential influence of the small non
force-free component, the arbitrary choice of a toroidal field seems to
be feasible.
For a fully self-consistent approach, the proper poloidal field
distribution has to be calculated from the current distribution
(see Eq.\,10).
We concentrate on a {\em quantitative} treatment of the local
equilibrium, rather than treating the star as a whole.

We assume a poloidal magnetic field structure of dipolar type, but
in general allow for a small deviation, either as a poloidal
non-dipolar component or an additional toroidal field.
Strong toroidal fields in white dwarfs are not implausible if 
generated by differential rotation (Muslimov et al. 1995).
It is not known whether toroidal field penetrates white dwarf surface
or remains contained in the interior.
So far, only poloidal magnetic fields were derived from the observations.
In the case of Ap stars the existence of large toroidal field on the
surface seems to be ruled out (see Landstreet 1987).

In summary, our basic assumption is that a distortion of the main (force-free)
dipolar field will induce a Lorentz force re-arranging the hydrostatic
equilibrium until a new steady state is reached.
In this sense we follow the approach of Hubbard \& Dearborn (1982) and
Stepie\'n (1978).
Meridional motion as possibly induced by toroidal fields
(see Mestel et al. 1988),
is neglected in our study.

In Sect.\,2 we recall the equations of magnetohydrostatic equilibrium
and discuss our basic assumptions.
In Sect.\,3 we investigate the magnetohydrostatic force balance
in one dimension.
In Sect.\,4 we discuss the two dimensional problem, including radiation
transfer in the diffusion limit and consider the method of finite elements.
The full problem could not yet be resolved numerically, however, we will
present preliminary solutions.
We finally summarize our results in Sect.\,5.

We will use both cylindrical $(R,\phi,Z)$ and spherical
coordinates $(r,\theta, \phi)$ depending on the symmetry of the individual
problem.
$(x,\phi,z)$ are cylindrical coordinates normalized to a stellar reference
radius $r_o = 10,000$\,km. 
We will use Gaussian (and astronomical) units in general, but add SI units
in parentheses whenever number values are derived.


\section{Radiative hydromagnetic equilibrium}


\subsection{Basic assumptions, notation}


Throughout the paper we will apply the following assumptions which 
simplify the general problem significantly.
(i) A static equilibrium,
(ii) ideal MHD (high conductivity),
(iii) axisymmetry, and
(iv) a grey atmosphere.
Further, (v) we will prescribe the magnetic field and electric current
distribution.

Real stellar atmospheres are neither static nor axisymmetric.
Ongoing convection leads to non-axisymmetric structures at least on the
medium spatial scale.
Steffen et al. (1995) show that hydrodynamic turbulence with convection
velocities of about $10$ km/s is present in the upper layers of DA white
dwarf atmospheres.
However, one may hypothesize that a strong magnetic field is able to
suppress the turbulent motion.
Moss (1979b) argues that in the case of main sequence stellar models the
magnetic field structure, which depends on parameters of field energy and
rotational energy, the main features follow already from radiative
equilibrium models without meridional circulation.
Wendell et al (1987) claim that in cooler white dwarfs the kinetic energy
density in convection zones is too small in order to affect
magnetic fields of strengths larger than 0.1 MG significantly.
There are many white dwarfs showing evidence for stellar oscillations, but no
example among the {\em magnetic} white dwarfs.
 
The field structure itself, derived from white dwarf spectra show that
the dominant field component is in general of dipolar structure,
but may also contain non-axisymmetric components leading to a total field
distribution offset from the stellar center (e.g. Putney \& Jordan 1995).
The field structure in the stellar interior is generally un-known. 
We assume that the dipolar field which is measured on the surface is a
global field and just continues throughout the atmospheric layer.
This is clearly questionable in respect to turbulent gas motions and the
high gas pressure in the deeper layers of the atmosphere.

We neglect any effects of the magnetic field on radiation transfer and 
assume a grey absorption coefficient in the form of Kramer's opacity.
The grey absorption assumption for white dwarf atmospheres is commonly used
also for small optical depths (see e.g. Litchfield \& King 1990).
It is known from both observed spectra and also spectral modeling that 
magnetic fields may alter the opacity (see Friedrich et al. 1994).
Since we are interested in the global structure of the white dwarf 
atmosphere, we consider such processes as less important.

The prescription of the magnetic field is equivalent to the assumption
of a stationary global field structure.
Even though the field pressure widely exceeds the gas pressure in the
outermost layers of the atmosphere, in the deeper layers the contrary
is true.
This region is applied as a boundary condition for the magnetic atmosphere.


\subsection{The assumption of ideal MHD}


In order to justify the assumption of ideal MHD for the local force-balance,
we compare the time scale for 
magnetic diffusion, 
$\tau_{\rm diff} = (4\pi /c^2) \sigma_c R^2$ 
with the dynamical (Alfv\'en) time scale, 
$\tau_{\rm dyn} = (R/B_{\star}) \sqrt{4\pi \rho}$.
The (Spitzer) conductivity is about
$\sigma_c \simeq 10^6 T^{3/2}$ (see e.g. Wedell et al. 1987), 
where $T$ is the gas temperature.
A first order estimate of temperature and density can be obtained if
we apply a standard white dwarf model with a grey atmosphere and
a polytropic gas law (e.g. Shapiro \& Teukolsky 1983).
In this case,
$T = T_{\star} \left(\rho/\rho_{\star}\right)^{\gamma -1}$, 
where
$T_{\star}$ and $\rho_{\star}$ are the critical values at the 
``bottom'' of the atmosphere where degeneracy sets in. 
Choosing $T_{\star}= 10^6$ and $\rho_{\star}\la 10^{-3}$ 
(Shapiro \& Teukolski 1983),
we derive 
\begin{equation}
\frac{\tau_{\rm diff}}{\tau_{\rm dyn}} \simeq
     10^{11}\left(\frac{T}{T_{\star}}\right)^{\frac{3}{4}}
     \left(\frac{T_{\star}}{10^6\,{\rm K}}\right)^{\frac{3}{2}}
     \left(\frac{\rho_{\star}}{10^{-3}{\rm g\,cm^{\!-3}}}\right)^{-\frac{1}{2}} 
     \left(\frac{R_{\star}}{10^9 {\rm cm}}\right)
     \left(\frac{B_{\star}}{10^8 {\rm G}}\right).
\end{equation}
This indicates that magnetic diffusion can indeed be neglected for the
local force-balance, especially for hot white dwarfs.
However, as it was shown in the literature (e.g. Landstreet 1987, Moss 1984)
they must not be neglected for the long time magnetic field evolution
of the star.


\subsection{Description of magnetic field}


Under the assumptions of stationarity and axisymmetry a magnetic flux function
$\Psi (R,Z) $ can be defined,
\begin{equation} 
 \Psi = \frac {1}{2 \pi} \int {\vec {B}}_{\rm P} \cdot d{\vec{A}} ,\quad\quad
 R{\vec{B}}_{\rm P} = \nabla \Psi \times  {\vec{e}}_{\phi }\,.
\end{equation}
In the magnetostatic approach, the poloidal electric current flows parallel
to the poloidal magnetic field,
\begin{equation}
I = I(\Psi ) = \int {\vec{j}}_{\rm P} \cdot d{\vec{A}} = 
\frac {c}{2} RB_{\phi}\,.
\end{equation}
This is similar to the case of non-static but stationary, force-free MHD.

\subsection{Radiation transfer}

The radiative equilibrium is described in the diffusive approximation,
\begin{equation} 
\nabla \cdot \frac {1}{\kappa\rho} \nabla T^4 = {\rm source\,\,term}\,,
\end{equation}
with a temperature $T$, and a mass density $\rho$.
This assumption is appropriate for optical depths $\tau > 1$.
We assume a 'grey' absorption coefficient in the form of Kramer's opacity,
\begin{equation} 
\kappa (\rho ,T) = \kappa_0 \rho T^{-3.5}
= {\kappa}_0 \frac{\mu}{\Re} P T^{-4.5}
\end{equation}
where $\kappa_0 = 4.34\times 10^{22}{\rm m^2 kg^{-1}}$, and
$\Re $ is the gas constant, $\mu$ the molecular weight of the gas particles,
and $P$ the gas pressure.
With Eq.\,(5) and the perfect gas law,
$P = \Re \rho T/\mu$,
the diffusion equation (4) can be rewritten in terms of
$P$ and $T$,
\begin{equation} 
\frac {1}{\kappa_0}\frac {4}{9.5}(\frac {\Re}{\mu})^2\,
\nabla \cdot \frac {1}{P^2} \nabla T^{9.5} = {\rm source\,\,term}.
\end{equation}

\subsection{Magnetohydrostatic equilibrium}

The static, hydromagnetic force equilibrium is described by
\begin{equation}
\frac {1}{\rho} \nabla (P + \pr) =
\frac {1}{4\pi \rho} (\nabla \times {\vec B}) \times {\vec B} -
\nabla \Phi .
\end{equation}
$\pr $ denotes the radiation pressure. 
Rotational effects are negligible,
\begin{equation}
\frac{\Omega^2R}{G\ms/R^2} \simeq 3\times 10^{-6}\,
\left(\frac{\ps}{1^{\rm h}}\right)^{-2}
\left(\frac{R}{r_o}\right)^3 
\left(\frac{\ms}{\msun}\right)^{-1}\,,
\end{equation}
for a stellar rotational period $\ps$ of the order of hours
($\Omega$ is the rotational frequency).
This is in contrast to e.g. Ap stars with much more extended atmospheres,
where the stellar rotation becomes important.
Thus, since it is the rotation that leads to meridional circulation flows, 
we can also neglect poloidal velocities.

The magnetic term in Eq.\,(7) can be rewritten in terms of
the flux function $\Psi $ and the poloidal current $I(\Psi)$,
\begin{equation}
(\nabla \times {\vec B}) \times {\vec B} = 
- \nabla \Psi \left( \nabla \cdot \frac {1}{R^2}\nabla \Psi 
+ \frac{2}{c^2 R^2}\left(I(\Psi)^2\right)'\right)\!,
\end{equation}
where the prime denotes the derivative $d/d\Psi$.
In axisymmetry, generally
\begin{equation}
R\,\nabla \cdot \frac {1}{R^2}\nabla \Psi = \frac{4\,\pi}{c} j_{\phi}\,,
\end{equation}
and the magnetic term can also be understood in terms of toroidal
and poloidal currents.
Equation (9) evidently shows that the magnetic field affects the 
force-balance only {\em across} the field (i.e. anti-parallel to 
$\nabla \Psi$).
Depending on the field geometry, this means that the strength of the 
$r$ and $\theta$-derivatives varies largely with stellar latitude.
In a dipolar-like distribution the $r$-derivative are small near the
pole and large near the equator.

As can be seen immediately from Eq.\,(9), and as it was pointed out 
by many authors before (e.g. Stepie\'n 1978), 
force-free field configurations do {\em not} contribute to the
pressure equilibrium.
In this case,
\begin{equation}
\nabla \cdot \frac {1}{R^2}\nabla \Psi =
- \frac{1}{R^2}\frac{4}{c^2}I(\Psi)\,I'(\Psi)\,,
\end{equation}
and the magnetic terms vanish together.
In particular, for a pure dipolar field, $I(\Psi) \equiv 0$.
Thus, only poloidal field geometries deviating from a dipolar structure 
(or, more general, from any superposition of force-free multi poles),
or a toroidal field, may influence the hydrostatic equilibrium.
With the perfect gas law, Eq.\,(9), and Poisson's equation
\begin{equation}
\nabla \cdot \nabla \Phi = 4 \pi G \rho\,,
\end{equation}
we calculate the divergence of Eq.\,(7),
\begin{equation}
\nabla \cdot T\,\nabla ({\ln} P) = 
- 4 \pi G\,(\frac{\mu}{\Re})^2\,\frac{P}{T} \nonumber 
- \frac{4\Re T}{\mu} \frac{\pr}{P} 
\left( \frac{\nabla T \nabla P}{T P} 
- 4.5\,(\frac{|\nabla T|}{T})^2 \right) 
- \nabla \cdot \left( \frac{T \nabla \Psi}{4\pi P} 
\,\left( \nabla \cdot \frac {1}{R^2}\nabla \Psi 
+ \frac{4 I\,I'}{c^2 R^2}\right)\right) 
\end{equation}
The factor $4.5$ results from the power index in Kramer's opacity.
Except for small optical depths, radiation pressure is negligible.

\subsection{Radiative magnetohydrostatic equilibrium}

With a prescribed magnetic field structure $\Psi(R,Z)$
and poloidal current distribution $I(\Psi) \sim R B_{\phi}$,
Eqs.\,(6) and (14) represent a pair of coupled partial differential
equations of second order with the unknown functions $P(R,Z)$ and 
$T(R,Z)$. 
Since the magnetic field may change the size of the stellar 
atmosphere, the size of the integration domain has to be adapted
during the numerical computations.
The stellar surface is properly determined by the condition
\begin{equation} 
|\frac {4}{3}\frac{\sigma}{\kappa \rho}\nabla T^4| = \sigma T^4.
\end{equation}
With that, a self-consistent solution can be obtained in an iterative
process.


\begin{figure}
\resizebox{93mm}{!}
{\includegraphics[height=5cm]{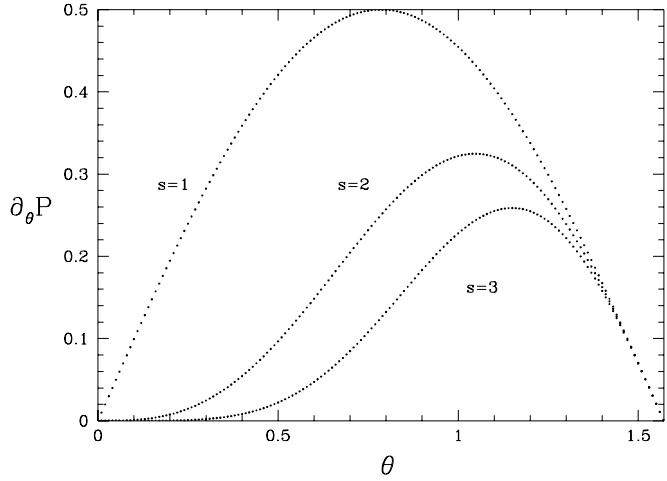}}
\hfill
\parbox[b]{63mm}{
\caption{
Normalized gas pressure gradient in $\theta$-direction, Eq.\,(17),
for different steepness $s$ of a model poloidal current distribution 
$I(\Psi) \sim {\Psi}^s$
}
\label{1}}
\end{figure}

\section{Magnetohydrodynamic equilibrium in one dimension}

\subsection{Surface temperature distribution}
As a preliminary estimate, we calculate the global change of temperature 
on a sphere with constant radius $r$.
In a dipolar field, the field strength varies from pole to equator by a
factor of 2 and thus the magnetic pressure $\pmag$ by a factor of 4, 
respectively.
We now introduce a factor $beta$ measuring the strength of the 
non force-free magnetic field component (see below). 
This factor should be rather low since the main field component is still
the dipolar field.
With the assumption that the total pressure (gas pressure plus magnetic 
pressure) is only a function of spherical radius, 
$\pt(r,\theta,\phi)\ =\pt(r)$,
the change of gas pressure by the magnetic pressure can be estimated.
For a polytropic gas law, $P \sim \rho^{\gamma}$, 
$T\sim P^{(\gamma-1)/\gamma}$,
the ratio between the equatorial and the polar temperature 
at a certain radius is
\begin{equation} 
\frac{T_{\rm eq}}{T_{\rm pole}} = 
\left(\frac{1 - P_{\rm eq} / P_{\rm tot}}
           {1-P_{\rm pole}/P_{\rm tot}}\right)^{\frac{\gamma-1}{\gamma}}
=\left(\frac{1-(4/9)\,\beta}{1-(16/9)\,\beta}\right)^{\frac{\gamma-1}
{\gamma}}\!\!\!.
\end{equation}
We define $\beta \equiv B^2_{\star}/(8\pi\pt)$, where 
$\bs = (B_{\rm eq} + B_{\rm pole})/2$.
We interpret $\beta $ as an ``effectiveness factor'' concerning the 
influence
of the magnetic pressure on the thermal pressure distribution,
having the following argumentation in mind.
Yet, magnetic tension, 
$\left({\vec{B}}\cdot\nabla\right){\vec{B}}$,
was not considered in our estimate.
For a dipolar field, magnetic tension exactly cancels the forces exerted
by the magnetic pressure gradient (a pure dipolar field is force-free).
The effectiveness factor $\beta$ can be understood as a measure of the non
force-free component of the global field.
These may be poloidal currents (or, equivalently, toroidal fields),
or a non force-free component of the poloidal field.
In Eq.\,(15) it is implicitly assumed that the non force-free components
depend linearly on the field pressure of the axisymmetric component.


\begin{figure}
\begin{minipage}[t]{53mm}
\resizebox{53mm}{!}
{\includegraphics[]{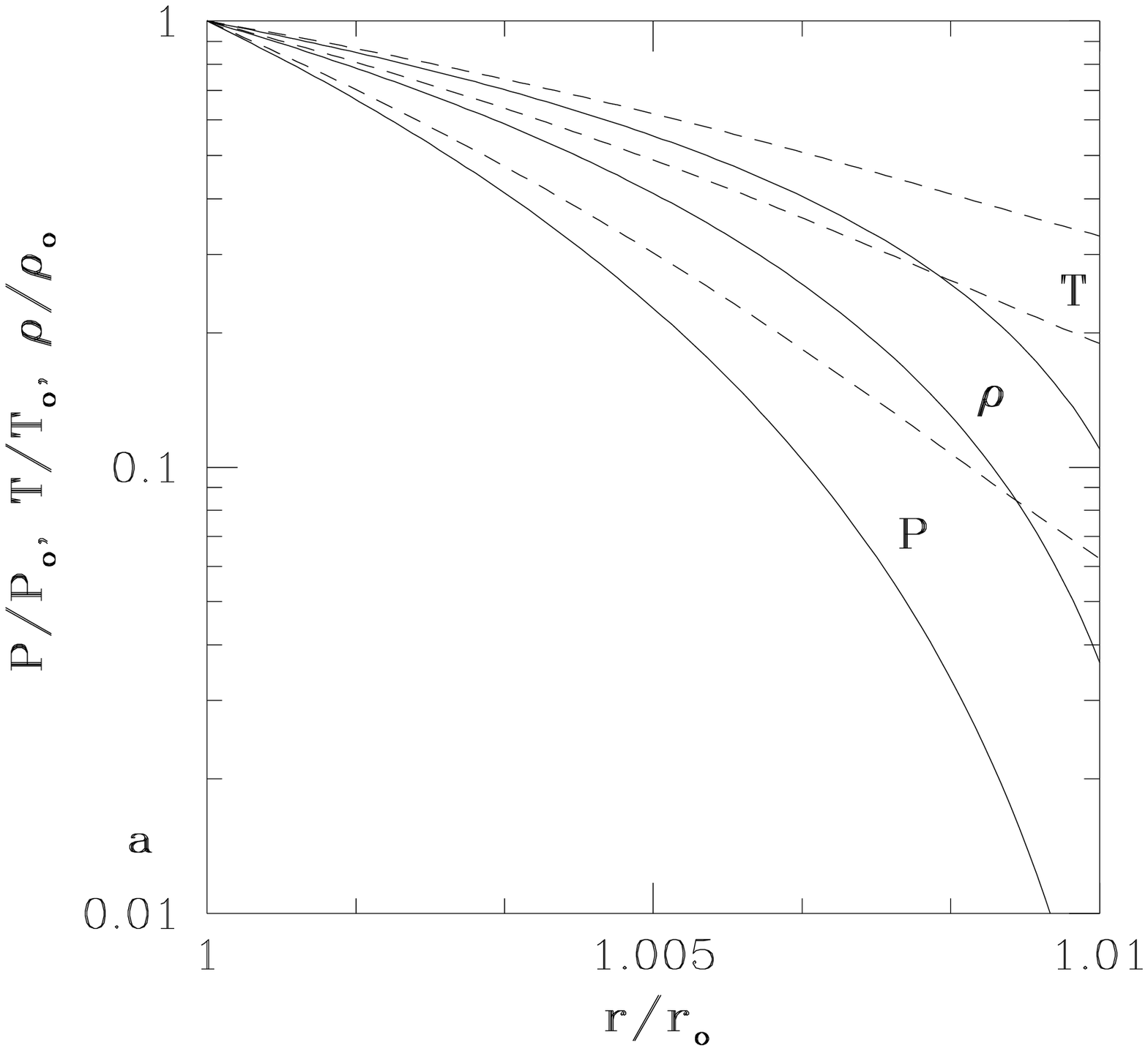}}
\end{minipage}
\begin{minipage}[t]{9mm}
\quad
\end{minipage}
\begin{minipage}[t]{53mm}
\resizebox{53mm}{!}
{\includegraphics[]{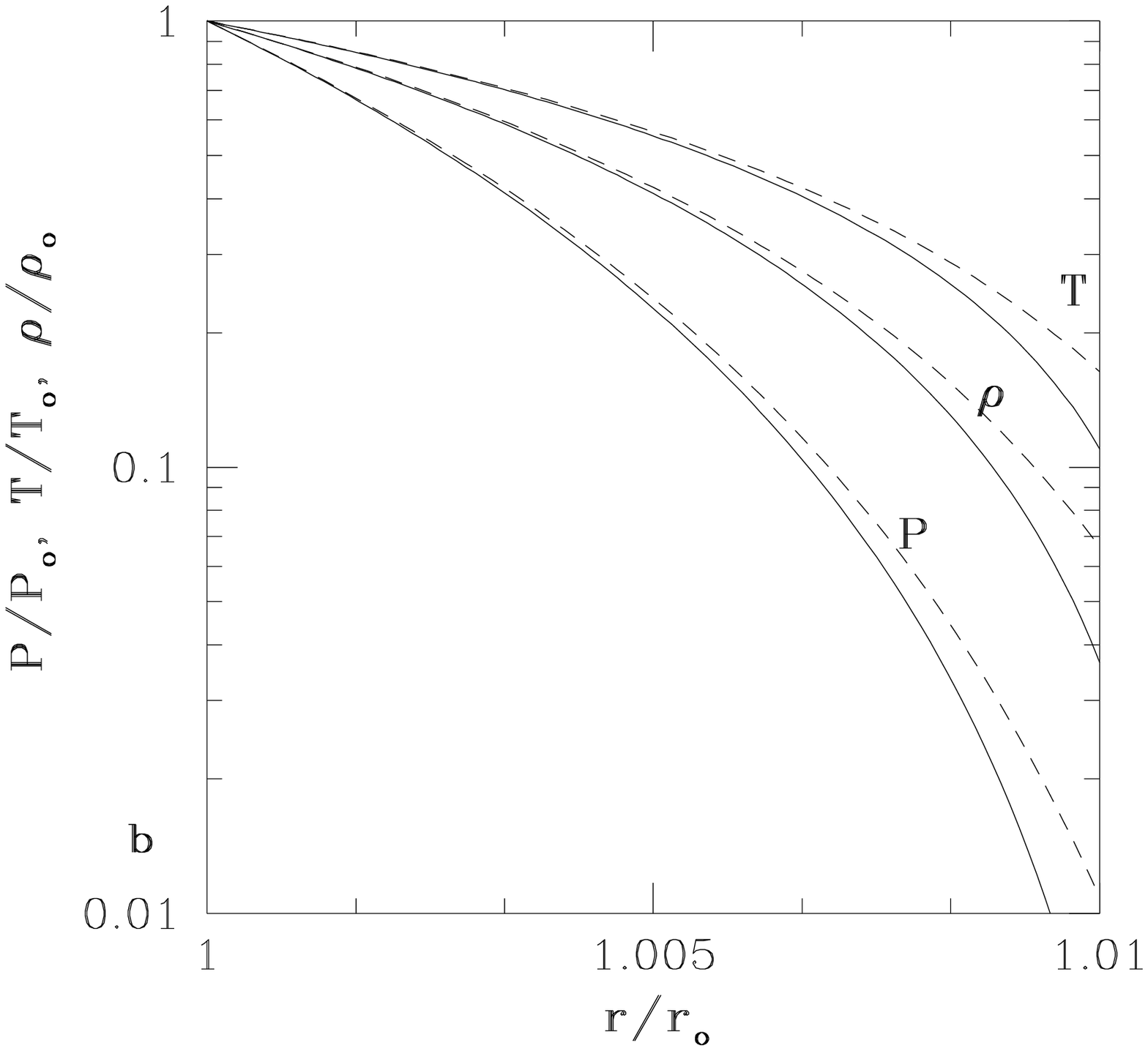}}
\end{minipage}

\begin{minipage}[t]{53mm}
\resizebox{53mm}{!}
{\includegraphics[]{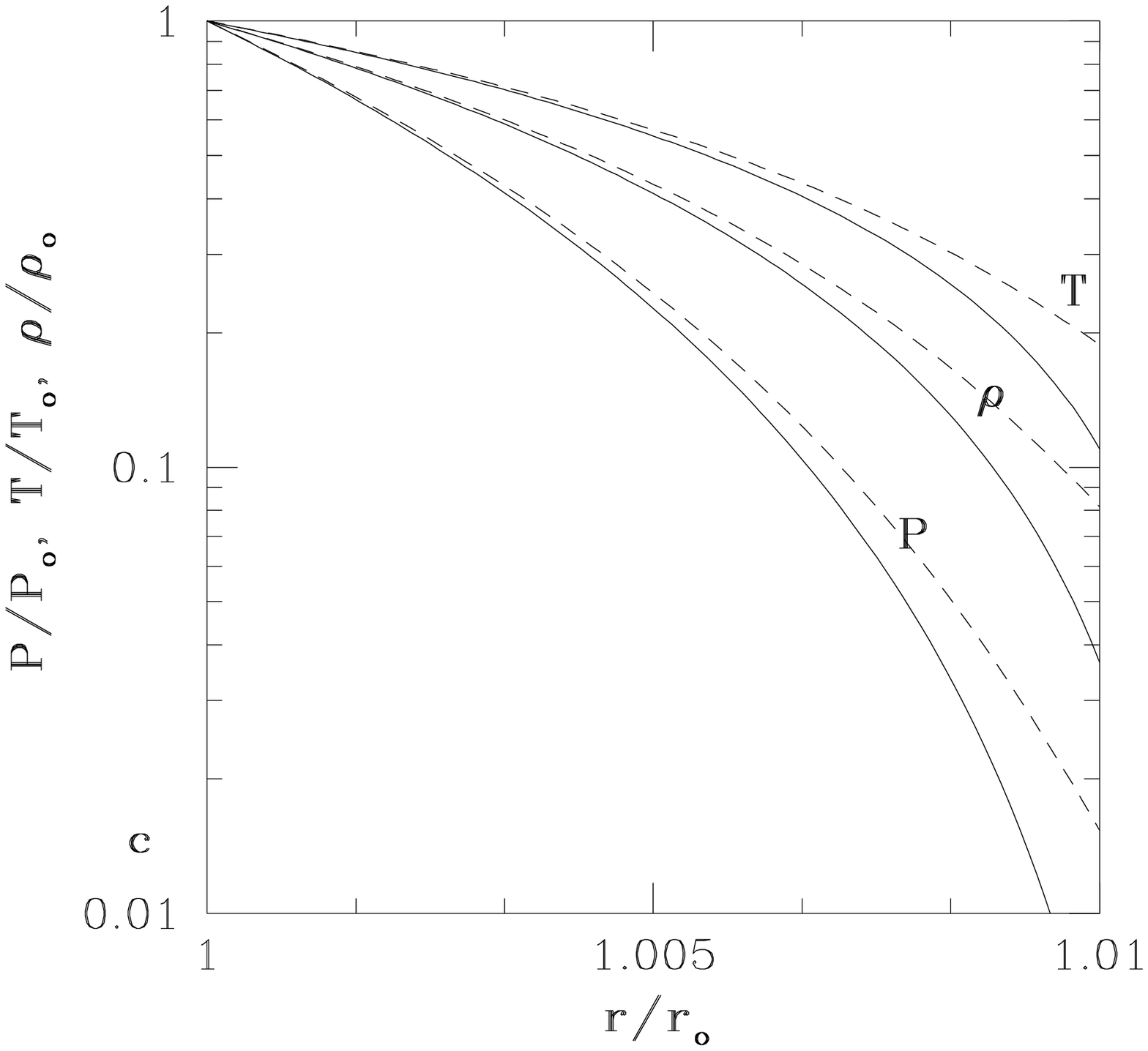}}
\end{minipage}
\begin{minipage}[t]{9mm}
\quad
\end{minipage}
\begin{minipage}[t]{53mm}
\resizebox{53mm}{!}
{\includegraphics[]{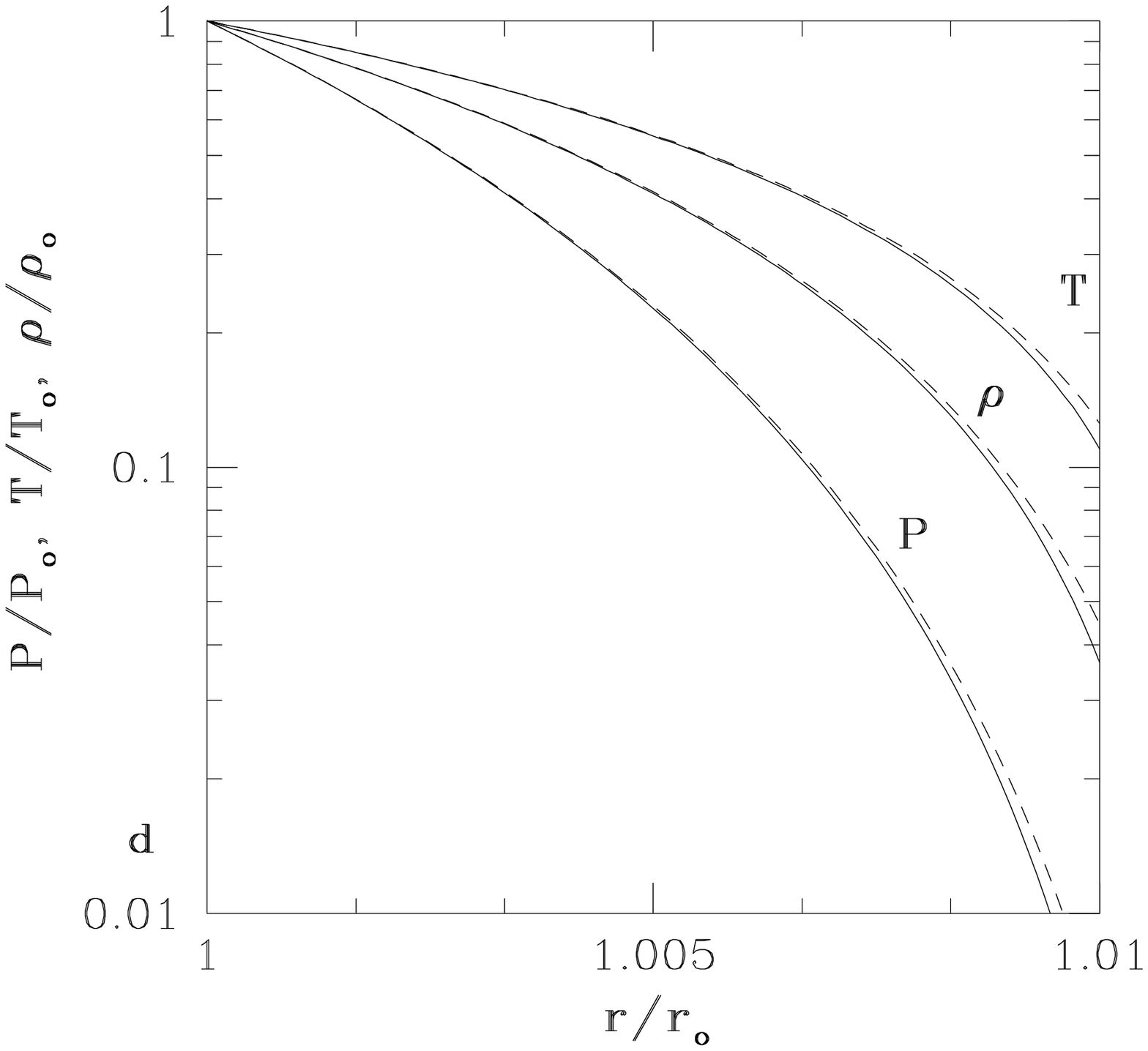}}
\end{minipage}

\begin{minipage}[t]{170mm}
\parbox[b]{165mm}{
\caption{Solution of the magnetohydrostatic equilibrium ignoring 
$\theta $-derivatives, and assuming a polytropic gas law. Gas pressure
$P$, mass density $\rho$, temperature $T$, normalized to their values
$P_o = 10^{14}{\rm dyne\,cm}^{-2} (10^{13}{\rm Pa})$, $T_o = 10^7$K
at the start point of the integration $r_o $. 
The polytropic index is $\gamma = 5/3$.
Solid lines denote the solution without magnetic field, dashed lines the
solution under influence of magnetic fields, respectively.
The latitude is $\theta = 0$ ({\bf a,b)}, and $\theta = (\pi/2)$ ({\bf c,d}) .
Field strength at $r_o$ is 100\,MG (10\,kT) ({\bf a,c}), 
and 40\,MG (4\,kT) ({\bf b,d}).
}
\end{minipage}
\label{2}}
\end{figure}

With $\gamma = (5/3)$ and choosing $\beta = 0.1$ the temperature variation
due to the global (non force-free) field is about $6\%$. 
With $\beta = 0.3$ the temperature variation increases to $20\%$.
Since the luminosity scales with $T^4$, the contribution from different
stellar latitudes might vary remarkably, even for small $\beta$.
$\beta = 0.1$ then refers to $13\%$ change in luminosity between 
polar and equatorial surface elements.
In this picture the magnetic poles might be considered as 'dark spots'
on the stellar surface.
The true temperature distribution depends on the distribution of the
electric current and non force-free field components.

From Eq.\,(7), it can be seen directly that a magnetic field, which is
not spherically symmetric, also implies an asymmetry in the gas pressure
distribution.
If Eq.\,(7) is re-written in spherical coordinates with a
gravitational potential $\Phi(r)=-(G M/r)$ and neglecting the radiation
pressure, 
the $\theta$-derivative of the pressure $\partial_{\theta}P$ is
\begin{equation}
\partial_{\theta}P(r,\theta) = 
- \frac{1}{4\pi}\frac{\partial_{\theta}\Psi} 
{r^2 \sin^2\theta}\,\frac{4}{c^2} I(\Psi)\,I'(\Psi)\,.
\end{equation}
For e.g. a dipolar field structure ($ \nabla \cdot (1/R^2)\nabla \Psi = 0$),
and a power law poloidal current distribution 
$I(\Psi) = I_{\rm max}\,{\Psi}^s$,
it simply follows
\begin{equation}
\partial_{\theta}P(r,\theta) \sim
\frac{1}{r^{2s+2}}\,(\sin\theta)^{4s-3}\cos\theta\,,
\end{equation}
which is shown in Fig.\,1.
The non spherically symmetric gas pressure distribution as a result of
poloidal currents is evident. 
In the case of dipolar fields and with a linear current distribution,
this effect scales with $r^{-4}$ and will influence different layers 
with different strength.

\subsection{An estimate of a critical radius for the expected magnetic
 effects} 
In the 1D limit, the differential equation for {\em hydro}static
equilibrium can be integrated analytically, applying diffusive
radiation transfer, a perfect gas law, and Kramer's
opacity approximation 
(see e.g. Shapiro \& Teukolsky 1983, Hansen \& Kawaler 1994),
revealing $P(r) \sim T^{4.25}(r)$. 
This approach of a grey atmosphere corresponds to a polytropic gas law
with $\gamma = (17/13)$.
The approximation $P(\rs)=0=T(\rs)$ gives
\begin{equation} 
T(r) = 7.55\times 10^8{\rm K}\,\left(\frac{\rs}{r} - 1\right)
\left(\frac{\ms}{M_{\sun}}\right) 
\left(\frac{\rs}{r_o}\right)^{-1}.
\end{equation}
With Eq.\,(18) and  $P(r) \sim T^{4.25}(r)$, a ``critical temperature'' can be
derived, where magnetic effects may become important, $\pmag \simeq P$.
The critical temperature is of the order of
\begin{equation} 
T_{\rm crit} = 3.6\times 10^5{\rm K}
\left(\frac{\bs}{500\,{\rm T}}\right)^{\!\!4/17}
\!\!\left(\frac{M}{\msun}\right)^{\!\!-2/17}
\!\!\left(\frac{L \mu}{10^{-3}\lsun}\right)^{\!\!2/17}\!.
\end{equation}
This corresponds to a critical optical depth in a grey atmosphere
(with $\kappa$ as in Eq.\,5), 
\begin{equation} 
\tau_{\rm crit} \simeq 1.2\times 10^6 
\left(\frac{\te}{10^4{\rm K}}\right)^{-4}\,,
\end{equation}
with otherwise the same parameters as in Eq.\,(19),
and a geometrical depth of $\simeq 10^{-4}\rs$.

The main conclusion from this estimate is that the magnetic field 
{\em potentially} may affect the hydrostatic structure in the 
atmosphere over a substantial region.
Clearly, the choice of 500\,T for the comparison with the gas pressure
overestimates the effect, since only the non force-free field component
may dynamically influence the matter.
Note that this rough estimate gives a result completely different
from the consideration of Jordan (1992) who calculated the
Lorentz force from electric currents rising in a conductive
atmosphere.
In our estimate, the one-dimensional limit is reasonable, since the
white dwarf atmosphere is very thin compared to the extension of the field,
the field variation being about $0.3\%$ along a radial scale of $0.001\,\rs$.

\subsection{Integration of the radial magnetohydrostatic equilibrium}
Here we integrate the radial component of the simplified
{\em magneto}hydrostatic equilibrium, 
$\nabla P = -\rho\nabla\Phi - \nabla\pmag$, in spherical coordinates.
The underlying assumption is that the $\theta$-derivatives are small
compared to the $r$-derivatives.
If we assume a dipolar-like distribution of the poloidal field
and a vanishing toroidal field,
the magnetohydrostatic equilibrium is described by
\begin{equation} 
\frac{dP}{dr} = -\frac {G M \rho_o}{r_o P_o}\frac{P^{1/\gamma}}{r^2}
+ \frac {B_o^2/8\pi}{P_o} \left(1-\frac{3}{4}\sin(\theta)^2\right)
\frac{24}{r^7}\,,
\end{equation}
where $r$, $P$, $B^2$, are normalized to their values at the
radius of the lower boundary of the non-degenerate atmosphere $r_o$
(note that here the radius $r_o$ is defined differently to the stellar
radius $\rs$ in Sect.\,3.2).
$T_o = (\Re/\mu)(P_o/\rho_o)$ is the temperature at this point.
The parameters of the stellar core (luminosity, mass, radius) are the same
for all solutions presented in this and the following subsection.
The strength of the r.h.s. terms of Eq.\,(21) scales with the ratio
of potential energy density to gas pressure at the lower surface, and
magnetic pressure to gas pressure, respectively.

Figure 2 shows solutions of Eq.\,(21) for different field strength $B_o$,
and for different stellar latitudes $\theta$
(We have used the mathematical software package {\em Mathcad} for
integrating Eq.\,(21)).
The general result is that the {\em magnetic pressure gradient} leads
to an {\em increase of scale height} of the atmosphere.
The gas pressure (and density and temperature) decreases faster in the
case without any magnetic field gradient.
This effect is larger near the axis, since the magnetic term in Eq.\,(21)
is strongest for $\theta =0$, although the $r$-gradients are larger
near the equator.
At the height of about $0.01\,r_o$ the gas pressure under influence of
magnetic field gradients is 15 times larger than without magnetic effects. 
For the gas density this factor is 6 (see Fig.\,2a).

In the case of very strong fields one observes an inversion of the density
profile at a certain radius, which is unphysical and shows the limits of
the current investigation.
The results presented in Fig.\,2 remain valid for any other global field
structure, 
for which the magnetic term is equivalent to the one in Eq.\,(21).
In such a case, $B_o$ can be understood as the non-dipolar component
of the dipolar-{\em like} structure, similar to Sect.\,3.1., and 
$B_o = \beta B_{\star}$. The numerical results will scale equivalently.

\subsection{Integration of the radial magnetohydrostatic equilibrium
including poloidal currents}
We now consider poloidal currents in the force-balance.
We assume a global dipolar poloidal field 
(thus, the first part in the parenthesis in Eq.\,(9) vanishes),
and prescribe, artificially, a poloidal current along the field lines
$I(\Psi)$.
Since the static magnetic forces direct {\em across the field} 
(see Eq.\,(9)), 
they will predominantly influence the hydrostatic equilibrium
at lower stellar latitudes,
in difference to the results of previous section. 
If we assume an underlying dipolar field, $\nabla \Psi$ is anti-parallel
to the radius vector.


\begin{figure}

\begin{minipage}[t]{53mm}
\resizebox{53mm}{!}
{\includegraphics[]{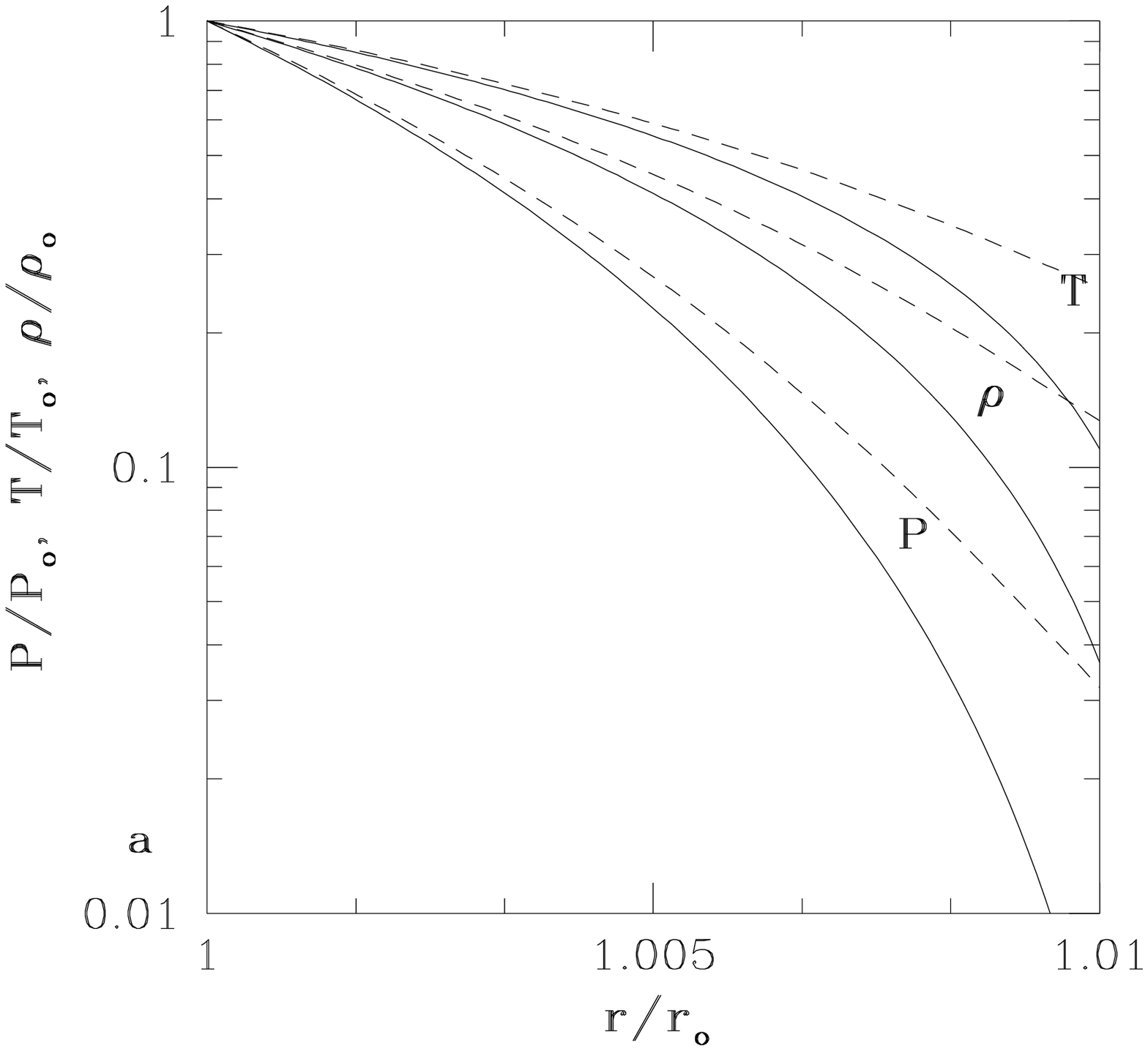}}
\end{minipage}
\begin{minipage}[t]{9mm}
\quad
\end{minipage}
\begin{minipage}[t]{53mm}
\resizebox{53mm}{!}
{\includegraphics[]{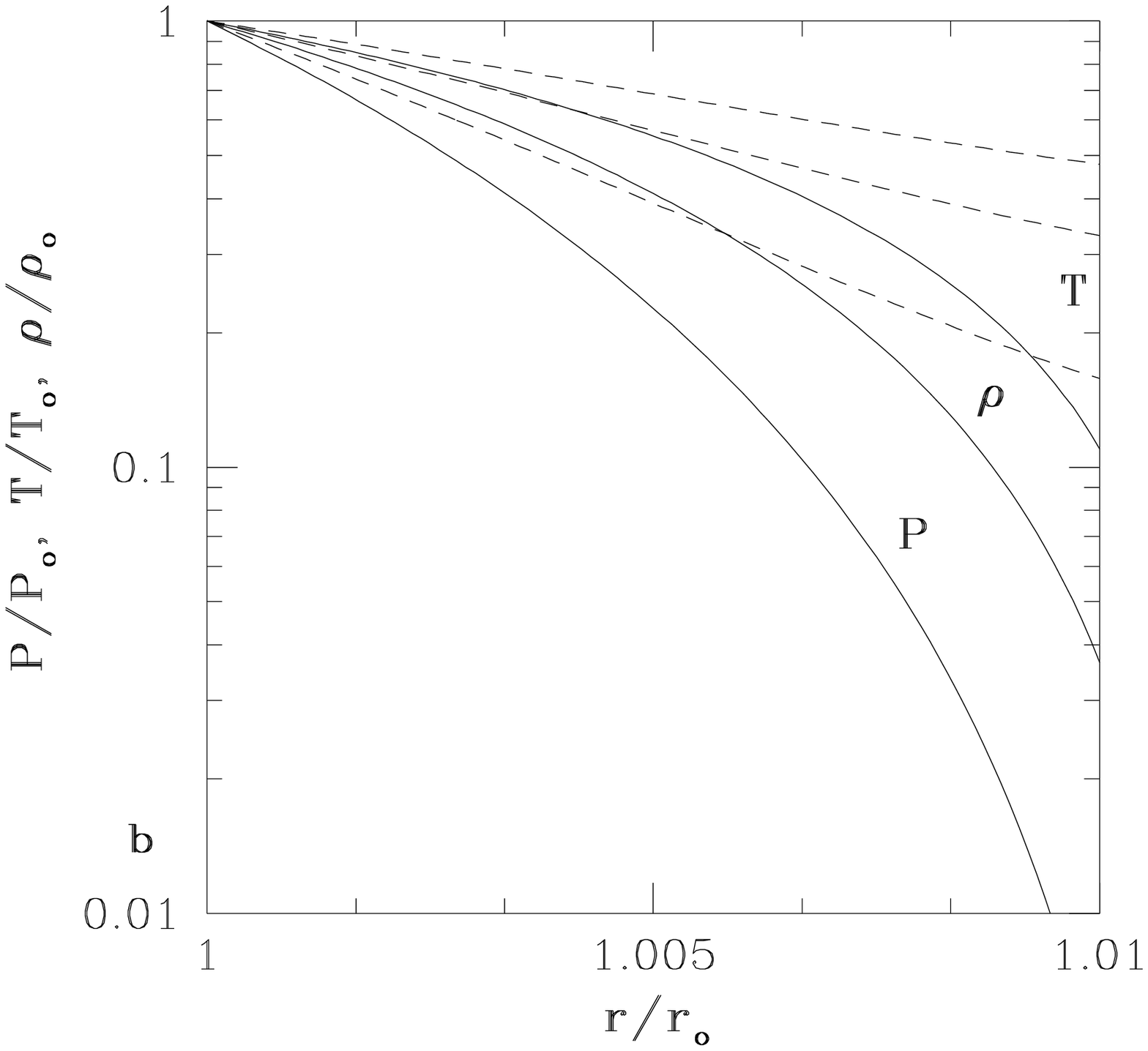}}
\end{minipage}

\begin{minipage}[t]{53mm}
\resizebox{53mm}{!}
{\includegraphics[]{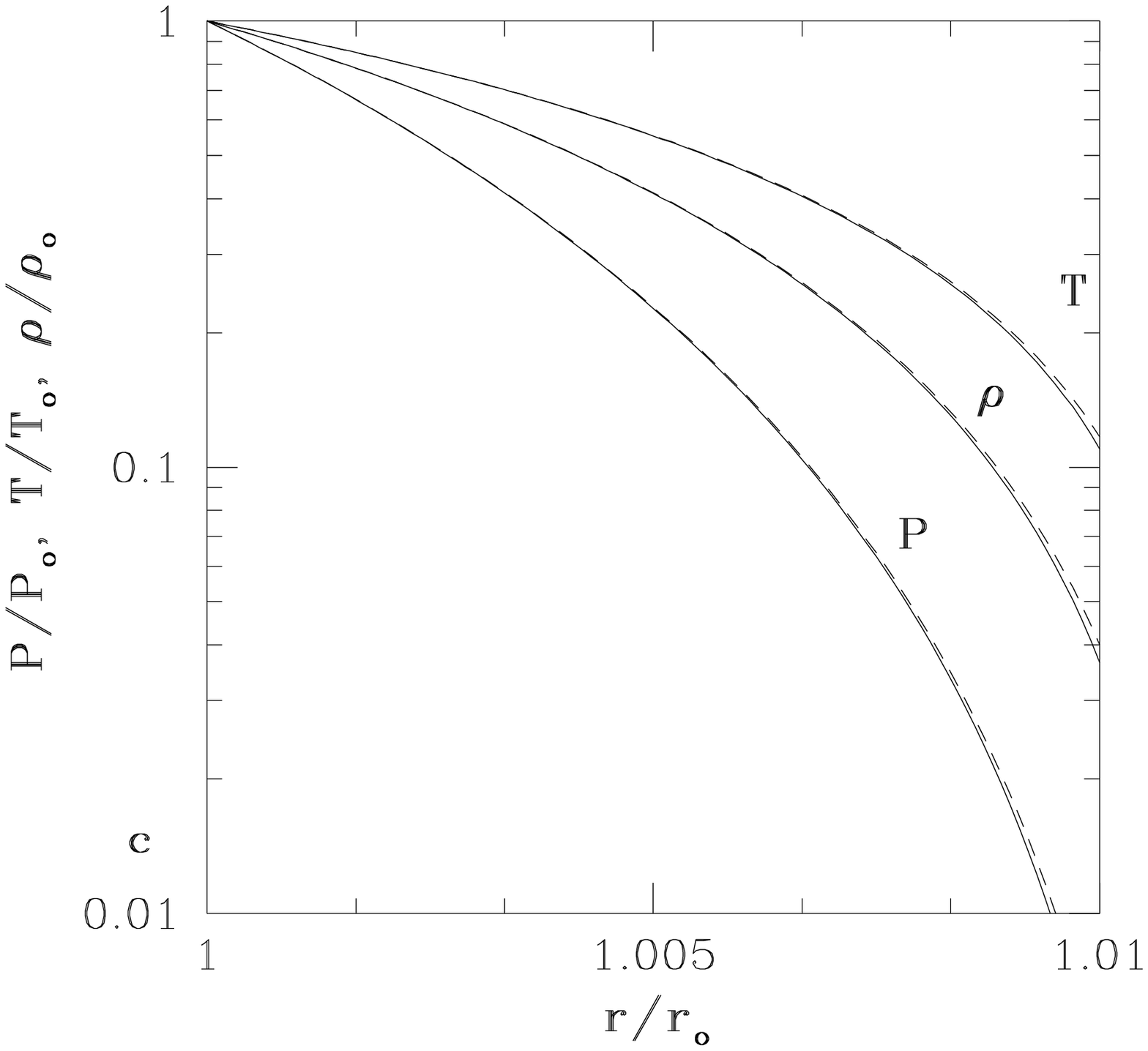}}
\end{minipage}
\begin{minipage}[t]{9mm}
\quad
\end{minipage}
\begin{minipage}[t]{53mm}
\resizebox{53mm}{!}
{\includegraphics[]{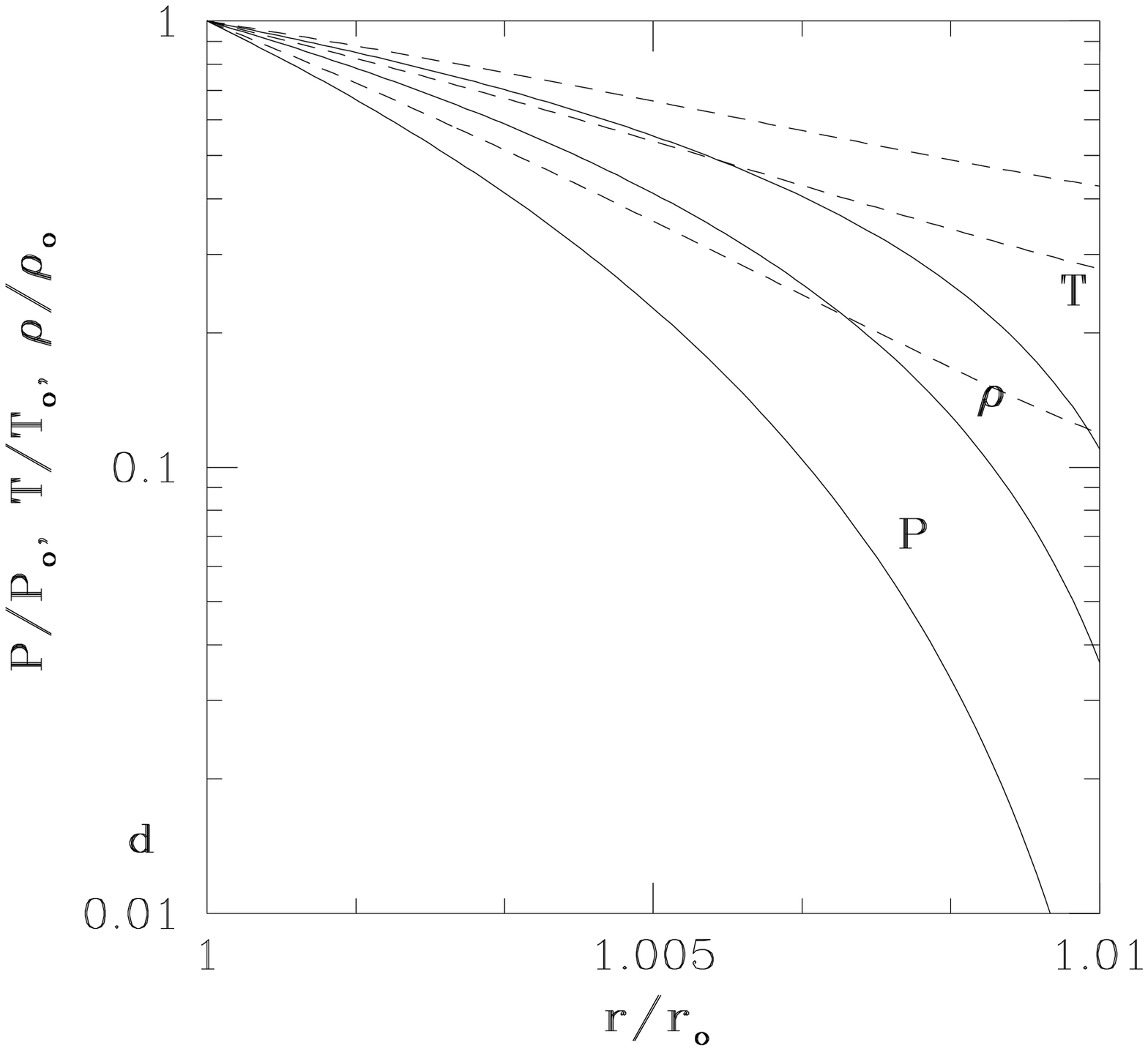}}
\end{minipage}

\begin{minipage}[t]{170mm}
\parbox[b]{165mm}{
\caption
{Solution of the magnetohydrostatic equilibrium including toroidal
fields / poloidal currents and assuming a polytropic gas law 
($\gamma = 5/3$).
$\theta $-derivatives are neglected. The latitude is $\theta = (\pi/2)$.
$P$, mass density $\rho$, temperature $T$, normalized to their values
$P_o = 10^{14}{\rm dyne\,cm}^{-2} (10^{13}{\rm Pa})$, $T_o = 10^7$K
at the start point of the integration $r_o $. 
Poloidal field strength at $r_o$ is $B_o = 100\,$MG (10\,kT).
Poloidal current distribution $I(\Psi) = A\,\Psi^n$.
Solid lines denote the solution without magnetic field, dashed lines the
solution under influence of magnetic fields, respectively.
Parameters: {\bf a} $\beta_c=0.01$, $n = 0.1$, $I_{\rm max}=5\,10^{12}$A,
{\bf b} $\beta_c=0.02$, $n = 0.1$, $I_{\rm max}=1\,10^{13}$A,
{\bf c} $\beta_c=0.01$, $n = 0.2$, $I_{\rm max}=5\,10^{12}$A,
{\bf d} $\beta_c=0.1$, $n = 0.2$, $I_{\rm max}=5\,10^{14}$A
}
\label{3}}
\end{minipage}
\end{figure}

A first constraint for a realistic current distribution is that the 
current must vanish along the symmetry axis, $I(\Psi = 0)$.
Therefore, in an axisymmetric magnetic field with dipolar geometry Lorentz
forces have no effect on the hydrostatic equilibrium along the symmetry axis.
The force-balance along the rotational axis is purely hydrostatic.
This is contrary to the scenario in Sect.\,3.1. and 3.3., where we assumed
that the magnetic effect on the hydrostatic force-balance depends solely 
on the field strength.

We chose the model poloidal current distribution as
$I(\Psi) = A\,\Psi^n$. 
The parameter $A$ is related to the maximum poloidal current $I_{\rm max}$
with $A = I_{\rm max} (1/B_o r_o^2)^n$. 
Then, for a dipolar field the radial component of the force-equilibrium is
\begin{equation} 
\frac{dP}{dr} = 
- \frac {G M \rho_o}{r_o P_o}\frac{P^{1/\gamma}}{r^2} 
- \frac {B_o^2}{8\pi} \frac{2 n {\beta_c}^2}{P_o r_o^{2 n -1}}
\frac{\sin(\theta)^{4 n -2}}{r^{4 n + 2}}\,.
\end{equation}
The parameter $\beta_c = (B_{\phi}/B_o)$ is defined as the ratio between the
poloidal and toroidal field strength, and $A \equiv (c/2) (B_o/r_o) \beta_c$.
It is equivalent to the parameter $\beta$ introduced in Sect.\,3.1.


\begin{figure}

\begin{minipage}[t]{73mm}
\bigskip
\bigskip
\bigskip
\bigskip

\end{minipage}

\begin{minipage}[b]{73mm}
\resizebox{73mm}{!}
{\includegraphics[]{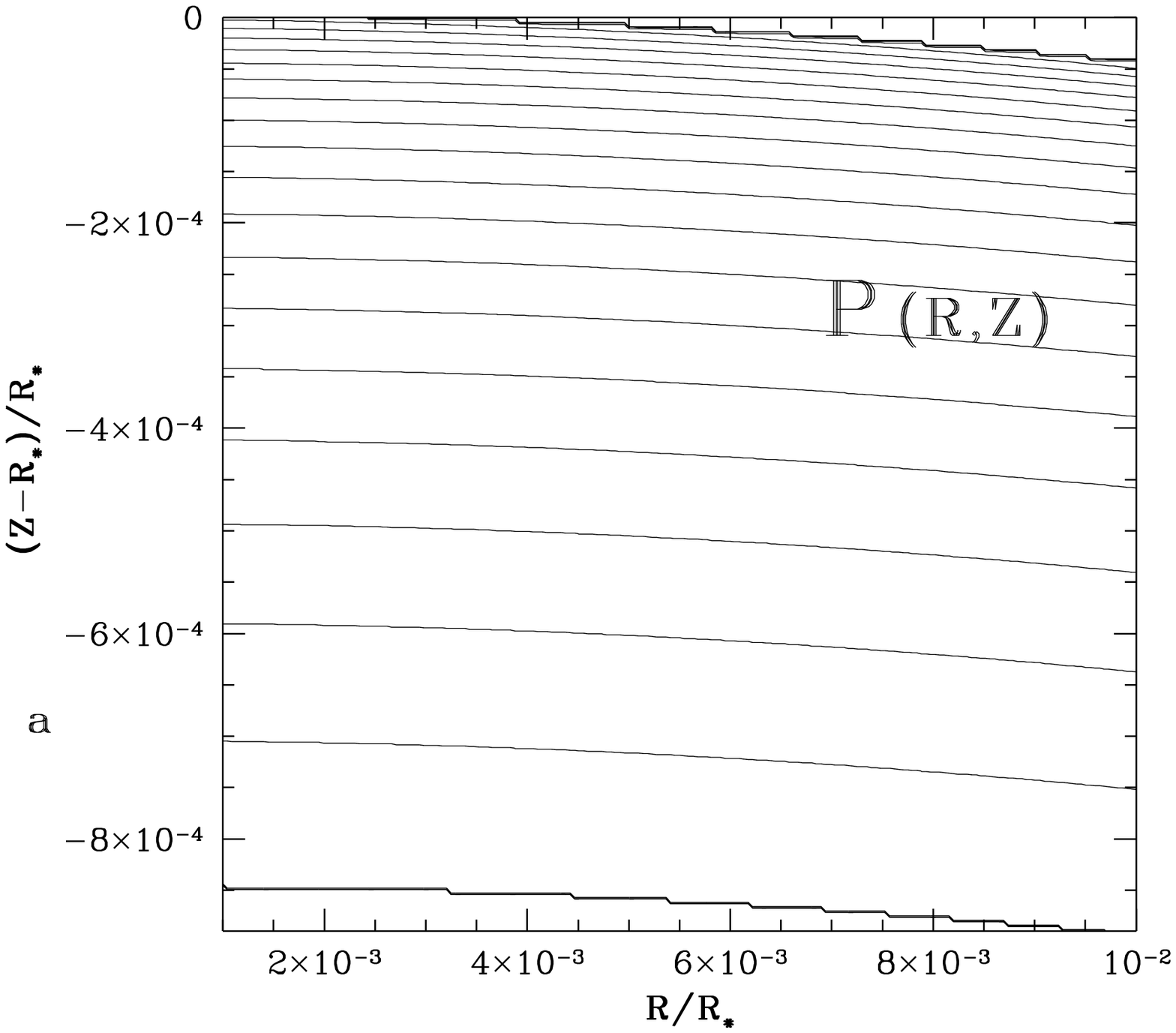}}
\par\vspace{0pt}
\end{minipage}
\begin{minipage}[b]{5mm}
\quad
\end{minipage}
\begin{minipage}[b]{73mm}
\resizebox{73mm}{!}
{\includegraphics[]{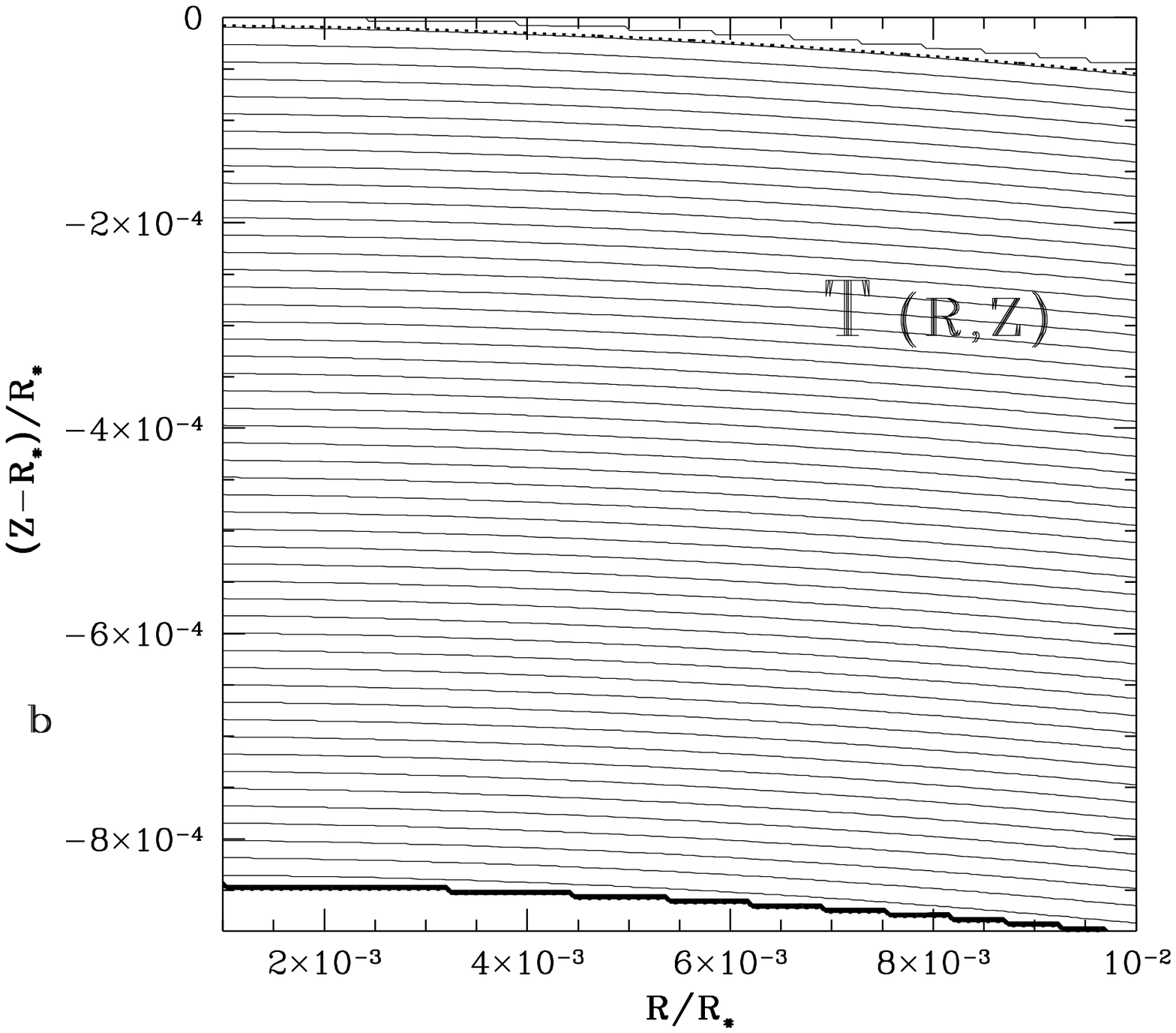}}
\par\vspace{0pt}
\end{minipage}

\begin{minipage}[t]{10mm}
\quad

\quad

\end{minipage}

\begin{minipage}[t]{10mm}
\quad
\end{minipage}
\begin{minipage}[b]{73mm}
\resizebox{73mm}{!}
{\includegraphics[]{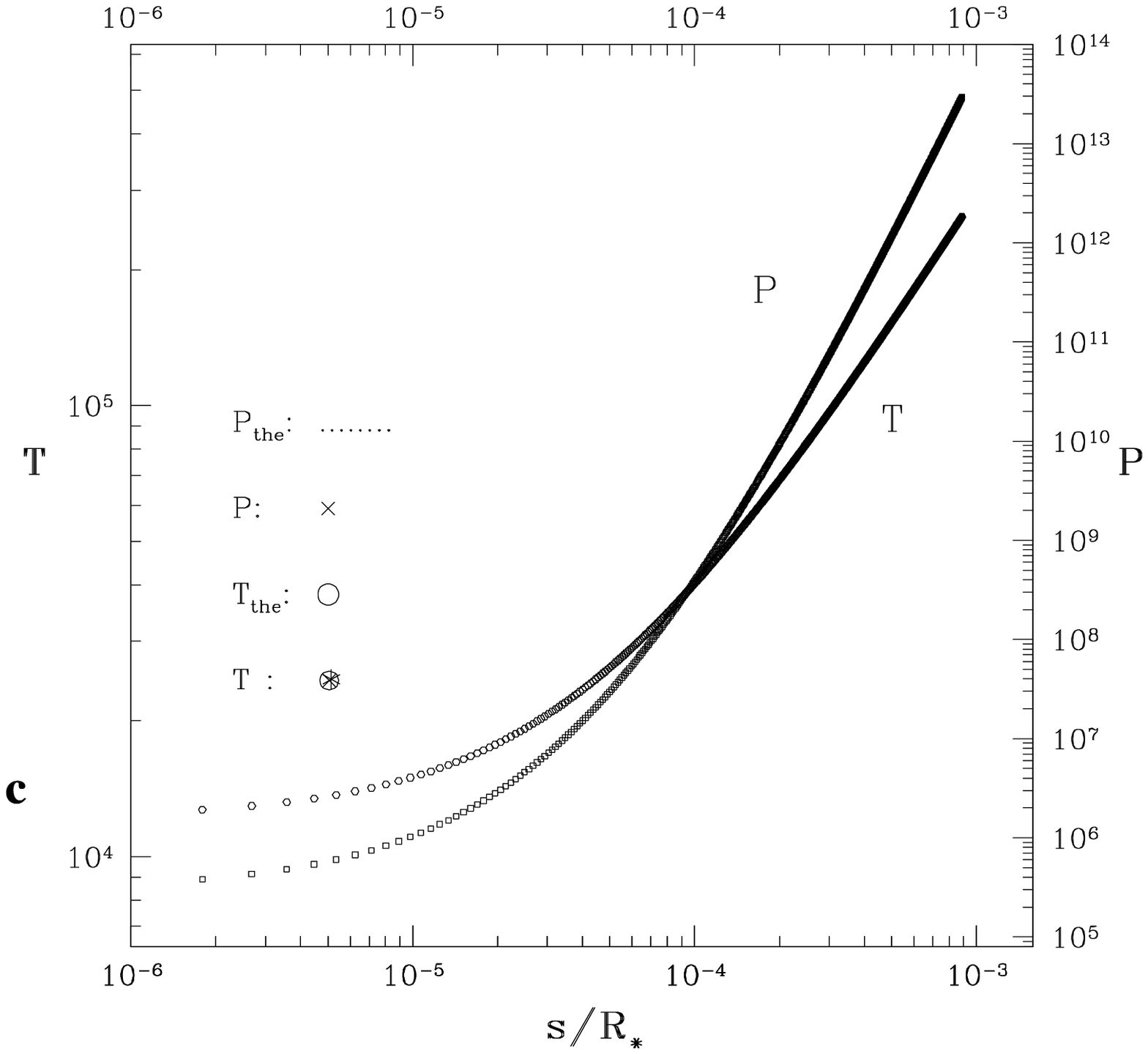}}
\par\vspace{0pt}
\end{minipage}
\begin{minipage}[t]{20mm}
\quad
\end{minipage}
\begin{minipage}[t]{63mm}
\parbox[b]{63mm}{
\par\vspace{0pt}
\caption{
Magnetohydrostatic equilibrium in the case of a non radiative
atmosphere with a force-free dipolar magnetic field of 
10\,MG (1 kT) field strength.
Isocontours of the two-dimensional {\bf a} pressure and {\bf b} temperature
distribution.
Contour levels: 
$P=\exp(n)\,{\rm dyne\,cm^{-2}}$ (0.1 Pa),
$n=5,30,1$; $T=n K$, $n= 10000, 300000, 5000$.
{\bf c} Pressure and temperature profile, $P$ and $T$,
along a radius vector near the right boundary in Figs.\,4a,b. 
$s$ is the normalized geometrical depth.
(Note that there are four curves, representing the theoretical 
(labeled with "the") and the numerical distribution, respectively)
}
\label{4}}
\end{minipage}
\end{figure}

Figure 3 shows the solution of Eq.\,(22) along the equatorial plane for
different strengths of the  poloidal current
(again, the {\em Mathcad} package was used for the integration procedure).
Similar to the results in Sect.\,3.3 the atmospheric scale height increases.
Again, for large currents an inversion of the density profile is indicated.
The results in Fig.\,3b,d, are close to this limit.

\section{The 2D radiative magnetohydrostatic equilibrium}
In Sect.\,2 we have formulated the equations of magnetohydrostatic equilibrium
and diffusive radiation transfer as partial differential equations of
second order with non-linear source terms.
In order to solve this type of equations the {\em method of finite elements} 
(see Camenzind 1987, Fendt et al. 1995 for further discussion and
tests of the code) is a suitable numerical approach.
The mesh of finite elements can be easily adapted to any shape of the
physical integration region.
Further, the finite element approach allows for a solution basically 
to any choice of boundary conditions.
In this section we discuss a first approach to numerical solutions of the
two-dimensional {\em magnetohydrostatic} equilibrium.

Besides the general advantage in using the method of finite elements,
the problem itself carries its own special difficulty.
On one hand we intend to solve the boundary value problem, 
in particular defined by the boundary of the stellar surface. 
But on the other hand, we do not know the position of that surface.
One of our aims is exactly to find the ``proper'' stellar surface under
influence of magnetic fields.

In particular, $\tau << 1$ and $\tau >> 1$ might be chosen as upper and
lower boundaries.
However, the diffusive radiation transfer assumption breaks down for small
optical depths, and the solution would not be self-consistent anymore.
The choice of the upper boundary to be close to the optical depth
$\tau \simeq 1$ is somehow in contradiction with the motivation to
calculate a ``new'' temperature distribution along the stellar
surface, since this is where the observed radiation originates.
Therefore, this region should not be too close to any prescribed
boundary condition.

For this preliminary investigation we chose the following approach.
Since at the lower boundary magnetic effects are negligible,
we use the analytical (hydrostatic) solution as 
a boundary condition for the gas pressure and temperature,
\begin{equation} 
P^2 = \left( \frac{2}{8.5}\frac{16\pi a c G \ms}{3\kappa_0 L_{\star}} 
\frac{\Re}{\mu} \right)
T^{8.5}\left( \frac{1-(T_o/T)^{8.5}}{1 - (P_o/P)^2}\right)\,,
\end{equation}
(e.g. Hansen \& Kawaler 1994).
With $P_o << P_l, T_o << T_l$ the gas pressure at the lower
boundary follows directly from the choice of the lower temperature.
Along the (spherically) radial boundaries we apply homogeneous Neumann
boundary conditions.
As the lower boundary we choose a certain geometrical depth and the
temperature/pressure according to Eq.\,(18) and (23).
Along the outer boundary we take a gas pressure of 
$P = 3\,10^5{\rm dyne\,cm}^{-2}$ ($3\,10^4{\rm Pa})$ 
adapted from the literature (Steffen et al. 1995) at $\tau = 0$.
The integration domain is about $10^{-3}\rs$ in the radial direction
and $\leq 10^{-2}\rs$ in a direction tangential to the stellar surface.

We use the following parameters.
A stellar mass of $\ms = 0.75 \msun$,
a luminosity of $2.5\times 10^{-4}\lsun$,
a radius $\rs = 10,000\,$km,
an effective temperature $\te = 12200\,$K,
an effective atomic mass $\mu = 1.4$.

\subsection{The test example -- the numerical solution with 
a polytropic gas law and force-free magnetic field}

If we neglect the radiative terms in Eq.\,(13), this equation 
allows for a investigation of electromagnetic effects on the
hydrostatic equilibrium solely.
Further, the case of a vanishing field or that of a dipolar field
may serve as a test for the numerical code.
Then, a polytropic temperature law $T(P)$ instead of Eq.\,(6)
is used for the iterative temperature update.
Figure 4 shows such test solutions of our code for a white dwarf atmosphere
with a polytropic index $\gamma = 6/5$ and a purely dipolar field 
of 10\,MG (1\,kT). 
The non-magnetic case gives the same result.
As mentioned above, we prescribe homogeneous Neumann boundary
conditions along the (spherically) radial boundaries. 
The numerical solution perfectly matches the analytical result similar to 
Eq.\,(23) and (18) (see also Fig.\,4c).


\begin{figure}

\begin{minipage}[t]{73mm}
\bigskip
\bigskip
\bigskip
\bigskip

\end{minipage}

\begin{minipage}[t]{73mm}
\resizebox{73mm}{!}
{\includegraphics[]{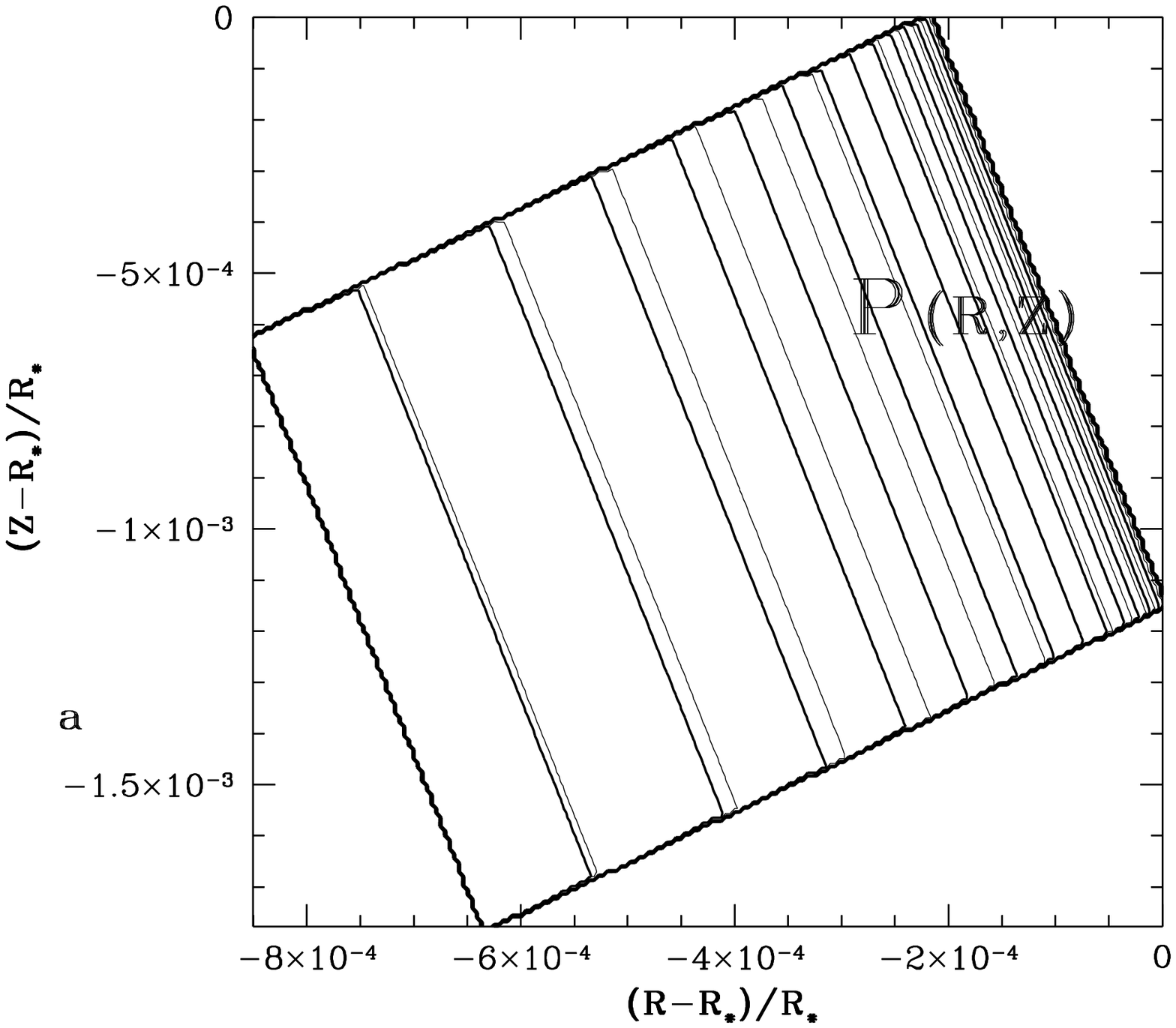}}
\par\vspace{0pt}
\end{minipage}
\begin{minipage}[t]{10mm}
\quad
\par\vspace{0pt}
\end{minipage}
\begin{minipage}[t]{73mm}
\resizebox{73mm}{!}
{\includegraphics[]{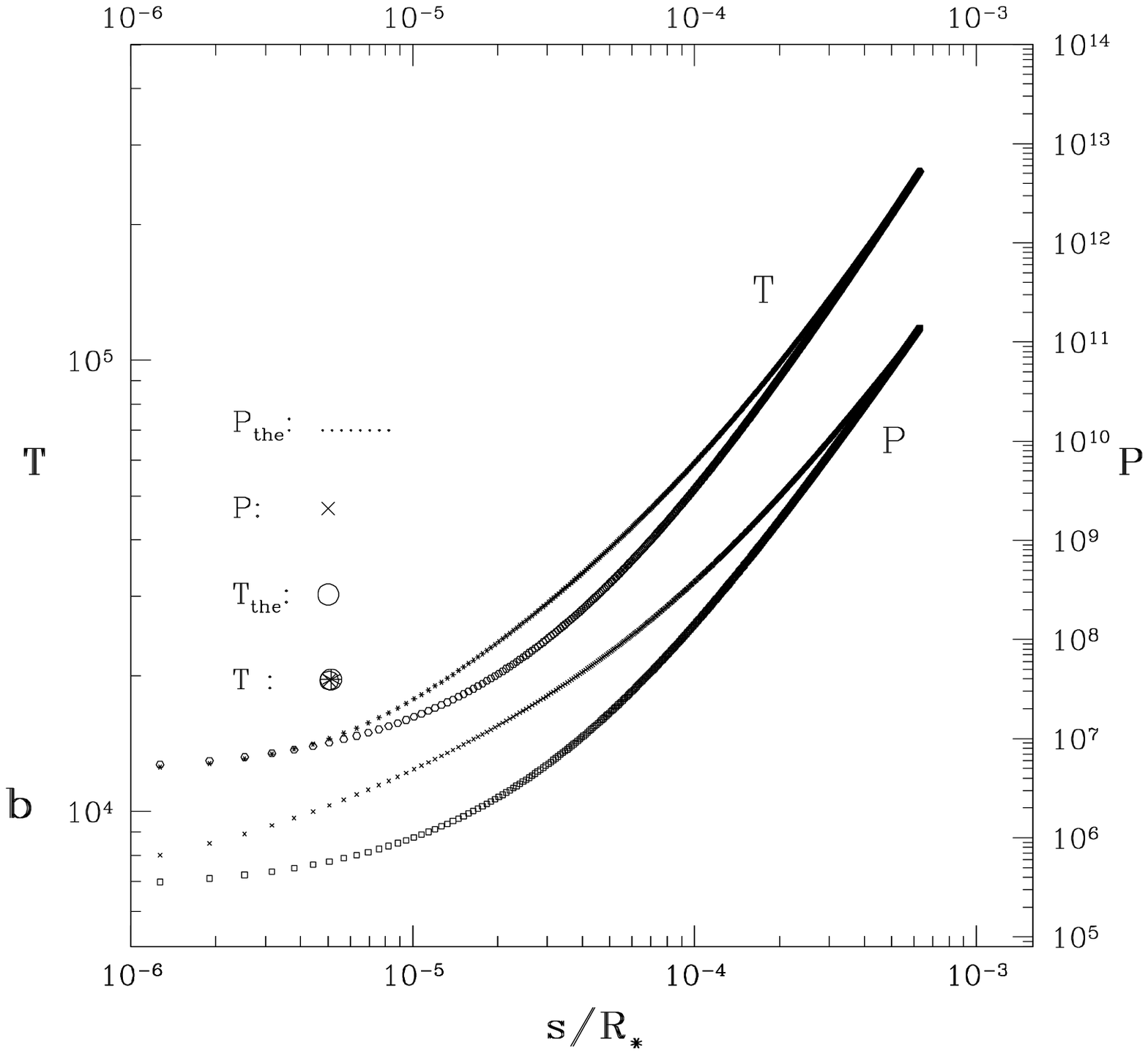}}
\par\vspace{0pt}
\end{minipage}

\begin{minipage}[t]{170mm}
\parbox[b]{165mm}{
\caption{
Magnetohydrostatic equilibrium for a magnetized white dwarf atmosphere
including diffusive radiation transport.
{\bf a} Isocontours of the two-dimensional pressure distribution
containing a force-free field {\it thick contours},
and also a toroidal field {\it thin contours}.
The integration region is located at 33\degr of stellar latitude.
The magnetic field consists of a 10\,MG (1 kT) dipolar field and 
a toroidal field (electric current) with $B_{\phi} \leq 1000 B_{\rm p}$.
Contour levels: 
$P=\exp(n)\,{\rm dyne\,cm^{-2}}$ (0.1 Pa),
$n=5,30,1$.
{\bf b} Distribution of pressure and temperature along a radius vector
in Fig.\,5a (the force-free solutions are labeled with "the").
}
\label{5}}
\end{minipage}

\end{figure}

\subsection{Non force-free magnetic field and diffusive radiation
transport}
The full solution of the two-dimensional hydromagnetic equilibrium
(Eq.(13)
including toroidal magnetic fields and
diffusive radiation transport (Eq.\,4)
is shown in Fig.\,5.
The calculation is done for an region located at a stellar latitude
of $\simeq 30\degr$
The upper boundary (the stellar surface) was fixed, i.e. we did
not consider Eq.\,(14).
We choose the same upper boundary condition as for the solution shown
in Fig.\,4.
The lower boundary, however, differs due to the treatment of diffusive
radiation transport.

The resulting two-dimensional gas pressure distribution is shown in
Fig.\,5a for the
the case of a force-free poloidal magnetic field and
the case where additionally a toroidal field is taken into account.
We have applied a current distribution
$I(\Psi) \sim (\Psi_{\rm max} - \Psi)(\Psi - \Psi_{\rm min})$, 
normalized in such a way that the maximum toroidal field is 
$B_{\phi} \leq 1000\,B_{\rm p}$.

For the solution considering the non force-free magnetic fields 
components, the gas pressure isocontours are shifted outwards
over large parts of the integration box.
The maximum increase of gas pressure is a factor of 8.
We hypothesize that, if we would have calculated the ``new'' position
of the stellar surface considering Eq.\,(4), also this surface would 
have moved outwards.
We propose that the increase of radius will be different for different
stellar latitudes.
It is only because of our choice of the upper boundary condition that
the pressure and temperature remain fixed at this boundary.
In this sense the solution close to the upper boundary layer is
preliminary, since the proper hydro{\em magnetic} upper boundary
condition for this boundary can only be found by an iterative procedure.
We defer this problem to a future paper.

In Fig.\,5b we plot the (spherically) radial pressure and temperature
profile close to the right boundary of the two-dimensional box in
Fig.\,5a.
Again, for comparison, 
the solution in the case of a force-free magnetic field is shown.
A dipolar of 10\,MG (1 kT) magnetic field leaves the hydrostatic
equilibrium unchanged (Figs.\,4c, 5b).
With the additional (non force-free) toroidal field component the gas
pressure and temperature {\em increase} in the intermediate region.

The relatively strong toroidal field strength is understandable, if we
consider that it is the {\em net} Lorentz force of the toroidal field, 
i.e. the difference between the magnetic pressure gradient directed outwards,
and the magnetic tension directed inwards, 
which changes the hydrostatic equilibrium.
For a different toroidal field distribution the field strength required
for the re-configuration of the hydrostatic equilibrium could be
considerably lower.

The question of the {\em observed} temperature distribution of the
inflated atmosphere cannot be answered yet.
So far, there seem to be two possibilities for the final state of the
hydromagnetic equilibrium, implying different observable effects. 
In one configuration the atmosphere inflates to such a structure
that the gas pressure isocontours and the temperature isocontours both
follow the same curve.
The other  possibility is that the surfaces of constant gas pressure
follow a curve {\em different} from that of the temperature.
This would be equivalent to a surface temperature variation along the 
stellar surface
corresponding to a re-distribution of radiation flux
in the direction perpendicular to the radius vector.

\section{Summary}

In this paper we investigated how the structure of white dwarf atmospheres
is affected by a global strong magnetic field.
Due to the complexity of the problem, and due to the lack of knowledge
about physical properties not yet delivered by the observations,
this question could only be treated under certain simplifying assumptions.
Our main assumption is the prescription of the magnetic field
distribution. 
We estimate, how the gas pressure and tenmperature changes if we
add a small non force-free field component to the large-scale force-free
magnetic field.

In the beginning we discussed some preliminary parameter estimates
assuming that the non force-free magnetic field component is proportional
to the total field strength, and its strength about 10\%.
With that the luminosity of a surface element may change by 20\% from 
polar to equatorial latitudes.

Then the magnetohydrostatic equilibrium was solved numerically in one
dimension along the radius vector.
We find clear evidence that the scale height of the white dwarf atmosphere
increases under the additional pressure of magnetic fields.
Depending on the field strength, the scale height of the atmosphere may
increase by a factor of 10. 
This is in agreement with spectropolarimetric observations 
(\"Ostreicher et al. 1992, Friedrich et al. 1994)
We point out that the variation of a non force-free field component will
also change the gas pressure across the surface and thus the temperature
and flux distribution.

The set of equations for the axisymmetric radiative magnetohydrostatic
equilibrium was re-formulated
into a pair of partial differential equations of second order.
These equations were solved with our code iteratively applying the method
of finite elements.
The two-dimensional distribution of the gas pressure and temperature 
in a box with a size of $0.001 \times 0.001$ stellar radii
located at $33\degr$ stellar latitude were presented.
As a result, the gas pressure and temperature levels are shifted to
higher altitudes.
Although the calculations were performed for a constant size of the
computational box our results gives a clear indication for 
an {\em expansion} of the white dwarf atmosphere under the influence
of the non-force-free magnetic field.
A more complete treatment would have to consider the proper determination 
of the true stellar surface.
However, our main result, i.e. the expanding atmosphere, is not
affected by that constraint.
Observational indictation for an increase of the scale height in magnetic
white dwarfs were found by Friedrich et al. (1994).

We hypothesize that the atmospheric expansion is different in amount for 
different stellar latitudes depending on the magnetic field distribution.
This would lead to a non-spherically distributed gas pressure and 
temperature profile in the atmosphere,
and, thus, to a variation of the luminosity of surface elements at
different latitudes..
Such a luminosity variation could finally be observed.

Barstow et al. (1995) concluded that for the hot and highly magnetic 
white dwarf 
RE\,J0317-853 the most plausible explanation for the modulation of optical
flux is due to the rotation of the star.
A non-equal global temperature distribution along the stellar surface due 
to magnetoelectric effects, as discussed in the present paper, together with
either an inclined rotational axis of the star or an inclined magnetic axis
would explain such an intensity variation.
Further observational developments, such as increased sensitivity for
time-resolved spectrophotometry and polarimetry in very large
telescopes, will provide a firmer experimental basis for verifying and
further developing these concepts.

\acknowledgements
This work was supported by the Swedish Natural Science Research Council
(NFR). C.F. thanks all members of Lund Observatory for their kind
hospitality.

\hangindent=0pt\hangafter=1
\begin{appendix}
\section{Numerical techniques: the finite element code}

Equations (13) and (6) are solved by means of the method of finite 
elements 
(for a more detailed discussion see Camenzind 1987, Fendt et al. 1995).
For this purpose the equations are multiplied by a test function $N$ 
(Galerkin ansatz) and integrated over the two-dimensional plasma domain $G$ 
applying Green's identity. 
In the case of Eq.\,(13) we end up with 
\begin{equation}
\int\limits_G R\,T \,
\nabla N \cdot \nabla (\ln P) \,dR\,dZ = 
 - \!\int\limits_G \!J\,N\,dR\,dZ
 + \!\int\limits_{\partial G} R\,T\, 
N \frac{\partial (\ln P)}{\partial n}\,dS,
\end{equation}
and with a similar result for Eq.\,(6).
$n$ denotes the unit vector perpendicular to the boundary $\partial G$,
and $J(R,Z)$ is the source term on the. r.h.s. of Eq.\,(13) or (6).

The integration domain $G$ is discretized in a set of isoparametric
curvilinear 8--node elements of the serendipity class (Schwarz 1991). 
Within each finite element the function $\ln P$ (or $T$) is expanded,
\begin{equation} 
\ln P(R,Z) = {\sum }_{i=1}^8 {\ln P}_i^{(e)}\; N_i(\zeta ,\eta ).
\end{equation}
${\ln P}_i^{(e)}$ denotes the logarithmic gas pressure at the nodal point
$i$ of the element $(e)$ and $(\zeta , \eta )$ are rectilinear coordinates 
on the normalized element.
The shape function $N_i$ is unity at each node $i$ and varies quadratically
with $\zeta $ or $\eta $ on the edge of the (Serendipity class of) element.

Following the Galerkin scheme we select the shape functions $N_i$ 
as test function and finally obtain a system of nonlinear equations 
for $\ln P$ at all nodal points,
\begin{equation} 
{\cal A}\;\ln P = \vec{b}(\ln P),
\end{equation}
with the integrals on each grid element 
\begin{equation} 
A_{ij}^{(e)} = \int_{G_e} R\,T 
\left( \,{\partial }_x N_i \;
 {\partial }_x N_j + {\partial }_z N_i \;{\partial }_z N_j 
\right) \,dx\, dz,
\quad {\rm and} \quad \quad
b_i^{(e)} = \int_{G_e} N_i\, J^{(e)}\; dx\, dz \;
+\; \int_{\partial G} R\,T \, N_i\, {\partial }_n (\ln P) \;ds ,
\end{equation}
and similarly for Eq.\,(6) and $T(R,Z)$.
Each component of Eq.\,(26) corresponds to the force equilibrium between
neighboring nodal points of each element. 
Inversion of matrix equation (26) yields the solution $(\ln P)$ for each
nodal point.
The expansion (25) provides the solution in any point $(\ln P(x,z))$. 

\end{appendix}

\refer
\aba
\rf{Barstow, M.A., Jordan, S., O'Donoghue, D., Burleigh, M.R.,
  Napiwotzki, R., Harrop-Allin, M.K., 1995: MNRAS, 277, 971}
\rf{ Camenzind, M., 1987: A\&A, 184, 341}
\rf{ Chanmugam, G., 1992: ARAA, 30, 143}
\rf{ Fendt Ch., Camenzind M., Appl S., 1995: A\&A, 300, 791}
\rf{ Friedrich S., \"Ostreicher R., Ruder H., Zeller G., 1994,
  A\&A, 282, 179}
\rf{ Hansen, C.J., Kawaler S.D., 1994, Stellar Interiors, Springer,
  Berlin}
\rf{ Hubbard E.N., Dearborn D.S.P., 1982: ApJ, 254, 196}
\rf{ Jordan, S., 1992: A\&A, 265, 570}
\rf{ Landstreet J.D., 1987: MNRAS, 225, 437}
\rf{ Landstreet, J.D., 1992: RvMAMex, 5, 437}
\rf{ Litchfield, S.J., King, A.R., 1990: MNRAS, 247, 200}
\rf{ Markiel, J.A., Thomas, J.H., van Horn, H.M., 1994: ApJ, 430, 834}
\rf{ Mestel, L., Moss, D.L., 1977: MNRAS, 178, 27}
\rf{ Mestel, L., Moss, D.L., 1983: MNRAS, 204, 575}
\rf{ Mestel, L., Moss, D.L., Tayler, R.J., 1988: MNRAS, 231, 873}
\rf{ Moss, D.L., 1975: MNRAS, 173, 141}
\rf{ Moss, D.L., 1979a: MNRAS, 186, 185}
\rf{ Moss, D.L., 1979b: MNRAS, 187, 601}
\rf{ Moss, D.L., 1984: MNRAS, 207, 107}
\rf{ Muslimov, A.G., van Horn, H.M., Wood, M.A., 1995: ApJ, 442, 758}
\rf{ \"Ostreicher, R., Seifert, W., Friedrich, S., Ruder, H., 
  Schaich, M., Wolf, D., Wunner, G., 1992: A\&A, 257, 353}
\rf{ Ostriker, J.P., Hartwick, F.D.A., 1968: ApJ, 153, 797}
\rf{ Putney, A., Jordan, S., 1995: ApJ, 449, 863}
\rf{ Schmidt, G.D., Smith, P.S., 1995: ApJ, 448, 305}
\rf{ Schmidt, G.D., West, S.C., Liebert, J., Green, R.F., 
  Stockman, H.S., 1986: ApJ, 309, 218}
\rf{ Schwarz, H.R., 1991: Methode der finiten Elemente, Teubner, 
  Stuttgart}
\rf{ Shapiro, S.L., Teukolsky, S.A., 1983: Black Holes, White Dwarfs,
  and Neutron Stars, Wiley-Interscience, New York}
\rf{Steffen, M., Ludwig, H.-G., Freytag, B., 1995: A\&A, 300, 473}
\rf{Stepie\'n, K., 1978: A\&A, 70, 509}
\rf{Stift, M.J., 1977: MNRAS, 178, 11 }
\rf{Stift, M.J., 1978: MNRAS, 183, 443}
\rf{Thomas, J.H., Markiel, J.A., van Horn, H.M., 1995: ApJ, 453, 403}
\rf{Wendell, C.E., van Horn, H.M., Sargent, D., 1987: ApJ, 313, 284}
\rf{Suh, I.-S., Mathews, G.J., 2000: ApJ, 530, 949}

\abe

\addresses 
\rf{Ch.~Fendt, Astrophysikalisches Institut Potsdam, An der Sternwarte 16,
DE-14482 Potsdam, Germany, cfendt@aip.de}
\rf{D.~Dravins, Lund Observatory, Box 43, SE-22100 Lund, Sweden,
dainis@astro.lu.se}

\end{document}